\documentclass{article}

\usepackage{tabularx}
\usepackage{algorithm}
\usepackage{graphicx}
\usepackage{authblk}
\usepackage{color}
\usepackage{fancyvrb}
\usepackage{amsmath,amssymb}
\usepackage{calc}
\usepackage{url}
\usepackage{doi}
\usepackage[authoryear,round]{natbib}

\author[1]{Pierre Masselot}
\author[1]{Devon Nenon}
\author[2]{Jacopo Vanoli}
\author[3]{Zaid Chalabi}
\author[1]{Antonio Gasparrini}

\affil[1]{Environment \& Health Modelling (EHM) Lab, 
      Department of Public Health, Environments \& Society, 
      London School of Hygiene \& Tropical Medicine, 
      London, 
      UK}

\affil[2]{Institute of Social and Preventive Medicine (ISPM), 
      University of Bern, 
      Bern, 
      Switzerland}

\affil[3]{UCL Institute for Environmental Design and Engineering, 
      The Bartlett School of Environment, Energy and Resources, 
      University College London, 
      London, 
      UK} 

\title{A constrained iteratively reweighted least squares framework for generalised linear models}

\begin{document}

%%% Title page
%%% -------------------------------------------------------------------------------
%%% Use command\maketitle to produce the title page.
\maketitle

%%% Abstract
%%% -------------------------------------------------------------------------------
\begin{abstract}
Many applications of generalised linear models (GLMs) can be improved by applying constraints that impose assumptions on the associations and improve consistency of the estimators. Yet, there are still barriers to the implementation and practical application of constrained GLMs. We present a general framework for fitting GLMs subject to linear constraints on the coefficients that offers original and interesting features. First, estimation is performed using constrained iteratively reweighted least-squares (CIRLS), offering fast and stable algorithms with excellent convergence performance. Second, the development includes advanced inferential procedures based on the truncated multivariate normal distribution and corrected degrees of freedom that account for the constrained nature of the estimation problem. Simulation studies indicate good inferential and computational properties, even in the case of slightly overconstrained models. Third, the proposed methods are fully implemented in the ‘cirls’ library for the R software, embedding constrained estimation in standard regression routines with simple usage and syntax. Two real-data case studies provide examples of applications for constrained dose-response estimation and compositional data analysis. The CIRLS framework and software offer a unified approach for various constrained estimation problems across a wide range of research areas.
\end{abstract}

%%% Main text
%%% INTRODUCTION
\section{Introduction}

Generalised linear models (GLM) are ubiquitous in statistics, being used in virtually every quantitative scientific field. GLMs model the association between a linear predictor and many different types of outcomes, including continuous, binary and count variables, through a link function that imposes a shape to this association. GLM modelling now consists of an established framework that also includes a comprehensive set of methods for inference, model selection, and diagnostics \citep{DobsonBarnett2018IntroductionGeneralizedLinear}.

There are situations in which it is useful to impose constraints on the coefficients of a regression model. Such constraints can represent specific assumptions or prior knowledge on the estimated associations, restricting unrealistic estimates or simply improving the fit. In this contribution, we focus generally on linear constraints, which already encompass a wide range of useful options. They allow simple models like sign-constrained least-squares \citep{Meinshausen2013SignconstrainedLeastSquares, SlawskiHein2013NonnegativeLeastSquares} as well as more complex models designed for high-dimensional data \citep{Tibshirani1996RegressionShrinkageSelection, GainesEtAl2018AlgorithmsFittingConstrained}. Linear constraints are also at the heart of non-parametric and spline-based shape-restricted regression \citep{MeyerWoodroofe2000DegreesFreedomShaperestricted, Meyer2013SemiparametricAdditiveConstrained}, as well as various types of regression models for compositional and other data relative to a reference \citep{AltenbuchingerEtAl2017ReferencePointInsensitive, AitchisonBacon-Shone1984LogContrastModels, Tsagris2025ConstrainedLeastSquares}.

There is a rich literature of applying linear constraints in a (non-generalised) linear regression context. A range of quadratic programs (QP) have been developed to fit such models \citep{Meyer2013SimpleNewAlgorithm, ZhouLange2013PathAlgorithmConstrained}, and inference procedures have been derived \citep{Davis1978BayesianAnalysisLinear, Geweke1996BayesianInferenceLinear}. However, the extension of such research to GLMs remains sparse. Some research provides frameworks for specific types of linear constraints, such as for shape-constrained splines \citep{Meyer2008InferenceUsingShaperestricted, PyaWood2015ShapeConstrainedAdditive}, ordinal regression \citep{DunsonNeelon2003BayesianInferenceOrderConstrained} or compositional regression \citep{LuEtAl2019GeneralizedLinearModels}. In the Bayesian setting, some research has also been conducted on the simulation of the posterior distribution of constrained coefficients for GLMs \citep{GhosalGhosh2022BayesianInferenceGeneralized}, but this requires the setting of a complex Gibbs sampling algorithm. Therefore, there is a lack of a general off-the-shelf framework with easily applicable software, putting practical barriers to the application of GLMs with linear constraints.

In this contribution, we propose a fast and flexible framework for GLMs subject to linear constraints on the coefficients, providing standardised procedures for their applications. The centrepiece of this framework is the constrained iteratively reweighted least squares (CIRLS) fitting algorithm, which exploits the speed and efficiency of quadratic programming \citep{GoldfarbIdnani1983NumericallyStableDual}. It quickly converges towards the solution, overcoming computational barriers to the practical application of constrained GLMs. Once a model is fitted, inference for the constrained coefficients can be performed through a truncated multivariate normal (TMVN) distribution \citep{Horrace2005ResultsMultivariateTruncated} from which confidence and prediction intervals can be obtained. We also account for the reduced effective complexity of constrained models, with a procedure to estimate their expected degrees of freedom that can typically be used with Mallows \(C_{p}\) or the Akaike Information Criterion \citep{Akaike1974NewLookStatistical, Mallows1973CommentsCP}. The full framework is implemented in the `cirls` package for the R software \citep{MasselotGasparrini2025CirlsConstrainedIteratively}, providing efficient routines and an easy-to-use syntax to fit various types of constrained models.

The remainder of the paper is organised as follows. A first section introduces the modelling framework and the CIRLS algorithm. It is followed by the related inferential framework, which includes the distribution of constrained coefficients and effective degrees of freedom. The next section reports the results of simulation studies, assessing the statistical and inferential properties of the framework. The CIRLS methodology is then illustrated in a real-world case study, while a final section provides a discussion.

%%%% METHODS
\section{The constrained iteratively reweighted least squares algorithm}
\subsection{Constrained generalised linear models}

Let \(y_i\), with \(i \in {1,...,n}\), be a real-valued observation from the exponential family of distributions, and \(\mathbf{x}_i\) a vector of \(p\) predictors. Our objective is to estimate the following GLM with a set of \(m\) linear constraints on the coefficients, i.e.:
\begin{equation} \label{eq:cGLM}
\begin{aligned}
g(\mu_i) = \mathbf{x}_i^{T}\boldsymbol{\beta} \\
\text{subject to} \quad & \mathbf{l} \le \mathbf{C}\boldsymbol{\beta} \le \mathbf{u}
\end{aligned}
\end{equation}
where \(\mu_{i}=\mathbb{E}(y_{i})\) and \(g(.)\) is a monotonic link function \citep{McCullaghNelder1989GeneralizedLinearModels}. \(\boldsymbol{\beta}\) is the \(p\)-dimensional vector of coefficients, subject to \(m\) linear constraints encoded by a \(m \times p\) constraint matrix \(\mathbf{C}\) with lower and upper bound \(m\)-dimensional vectors \(\mathbf{l}\) and \(\mathbf{u}\), respectively. The definition in Equation (\ref{eq:cGLM}) encompasses all types of linear constraints. These include, for instance, equality constraints when \(l_j = u_j\) for any constraint \(j\), as well as one-sided constraints by setting \(l_j=-\infty\) or \(u_j=\infty\). Supplementary Material A shows how \(\mathbf{C}\),  \(\mathbf{l}\) and \(\mathbf{u}\) are set-up for several common constraints.

% Algorithm
\subsection{Constrained iteratively reweighted least squares}

The unconstrained GLM is usually fitted with the iteratively reweighted least squares (IRLS) algorithm, where, at each iteration, the estimated \(\boldsymbol{\beta}\) is updated by minimising
%Equation
\begin{equation} \label{eq:IRLS}
\begin{split}
\hat{\boldsymbol{\beta}}^{[k+1]} & = \min_{\boldsymbol{\beta}} \sum_i w_i \left( z_i - \mathbf{x}_i \boldsymbol{\beta} \right)^2 \\
    & = \min_{\boldsymbol{\beta}}
       \left( \mathbf{z} - \mathbf{X}\boldsymbol{\beta} \right)^{T}
       \mathbf{W}
       \left( \mathbf{z} - \mathbf{X}\boldsymbol{\beta} \right)
\end{split}
\end{equation}
with \(\mathbf{X}\) \textbf{a} \(n \times p\) matrix that includes \(\mathbf{x}_{i}\) as its \(i^{th}\) row. The pseudo-response \(\mathbf{z}\) and pseudo-weight diagonal matrix \(\mathbf{W}\) are recomputed at each iteration depending on the current \({\widehat{\boldsymbol{\beta}}}^{\lbrack k\rbrack}\) and on the specific distribution of \(y_{i}\). Their full expression is provided in Supplementary Material B, and their derivation is covered elsewhere \citep{McCullaghNelder1989GeneralizedLinearModels}. The minimisation problem in Eq. (\ref{eq:IRLS}) is iterated until convergence, \emph{e.g.} until the decrease in deviance remains below a predefined small threshold.

In the extension to the constrained iteratively reweighted least squares (CIRLS) algorithm, we subject each step of the algorithm in Equation (\ref{eq:IRLS}) to the constraints of Equation (\ref{eq:cGLM}) so that each update \({\widehat{\boldsymbol{\beta}}}^{\lbrack k\rbrack}\) remains feasible. Rearranging the least-squares function of Equation (\ref{eq:IRLS}), we have the following constrained optimisation problem:
%Equation
\begin{equation} \label{eq:CIRLS}
\begin{split}
\hat{\boldsymbol{\beta}}^{[k+1]} & = \min_{\boldsymbol{\beta}}
       \left(
           \boldsymbol{\beta}^T \mathbf{X}^T \mathbf{W} \mathbf{X} \boldsymbol{\beta}
           - 2 \mathbf{z}^T \mathbf{W} \mathbf{X} \boldsymbol{\beta}
           + \mathbf{z}^T \mathbf{z}
       \right) \\
       \text{s.t.} & \quad \mathbf{l} \le \mathbf{C}\boldsymbol{\beta} \le \mathbf{u}
\end{split}
\end{equation}
The CIRLS step, as laid out in Equation (\ref{eq:CIRLS}), is a typical Quadratic Program (QP) for which many efficient algorithms exist \citep{BoydVandenberghe2004ConvexOptimization}. In the simulations and applications below, the QP in Equation (\ref{eq:CIRLS}) will be solved by a dual algorithm \citep{GoldfarbIdnani1983NumericallyStableDual}, but other equally efficient algorithms can be considered, such as the alternating direction method of multipliers \citep{StellatoEtAl2020OSQPOperatorSplitting}, or cone projection \citep{Meyer2013SimpleNewAlgorithm}.

Another way to view the CIRLS algorithm is as a Sequential Quadratic Program (SQP). These algorithms are useful for solving general nonlinear constrained optimisation problems and, under some conditions that are generally met in the specific case of constrained GLMs, tend to converge quickly \citep{BoggsTolle1995SequentialQuadraticProgramming}. The full CIRLS algorithm and how it relates to SQP is detailed in Supplementary Material B.

%%% Inference
\section{Inference}

When regression is fitted with constraints, the usual asymptotic theory providing distributions and inference for the coefficients is no longer valid \citep{Wets1991ConstrainedEstimationConsistency}. Indeed, the application of typical formulas results in probability distributions that can include unfeasible coefficient vectors, \emph{i.e.}, coefficients that violate the constraints in Equation (\ref{eq:cGLM}). In this section, we describe the inference for coefficients estimated by CIRLS and model selection for the constrained GLM. As detailed in Supplementary Material C, here we assume that \(n \geq p\) and \(m \leq p\). Supplementary Material C also provide an algorithm for constraint matrix reduction to ensure there is no redundant constraint in \(\mathbf{C}\) and \(m\) is the lowest it could be.  

\subsection{Distribution}

In unconstrained models, the estimated coefficient vector \(\widehat{\boldsymbol{\beta}}\) is asymptotically multivariate Gaussian. Truncating such a distribution has been a longstanding topic of analysis \citep{Tallis1965PlaneTruncationNormal} and the resulting TMVN is now well characterized \citep{Horrace2005ResultsMultivariateTruncated}. More specifically for constrained GLMs, it can be shown that the linearly transformed coefficient vector \(\mathbf{C}\widehat{\boldsymbol{\beta}}\) follows a TMVN \citep{Geweke1996BayesianInferenceLinear}, \emph{i.e.} we have that

%Equation
\begin{equation} \label{eq:TMVN}
\mathbf{C}\hat{\boldsymbol{\beta}} \sim \text{TMVN} \left(\mathbf{C}\boldsymbol{\beta}^*, \phi^*\,\mathbf{C}(\mathbf{X}^T\mathbf{W}^*\mathbf{X})^{-1}\mathbf{C}^T,
        \boldsymbol{l},
        \boldsymbol{u}
    \right)
\end{equation}
where \(\boldsymbol{\beta}^*\), \(\mathbf{W}^*\), and \(\phi^{*}\) are respectively the coefficient vector, weight matrix, and dispersion parameters from an \emph{unconstrained} model, \textit{i.e.} obtained through the classical IRLS algorithm in Equation (\ref{eq:IRLS}). Note that the distribution in Equation (\ref{eq:TMVN}) is also found from a Bayesian perspective using a typical uninformative prior multiplied by an indicator function that assigns null probability mass to unfeasible coefficient vectors \citep{Davis1978BayesianAnalysisLinear, Geweke1986ExactInferenceInequality, GhosalGhosh2022BayesianInferenceGeneralized}.

Some work has explored the properties of TMVN distributions \citep{Horrace2005ResultsMultivariateTruncated}, proposing formulas for the moment generating function \citep{Tallis1961MomentGeneratingFunction, Tallis1965PlaneTruncationNormal} and, from there, the first and second moments \citep{KanRobotti2017MomentsFoldedTruncated, ManjunathWilhelm2021MomentsCalculationDoubly}. However, these formulas do not allow the computation of various inferential quantities, \emph{e.g.} confidence intervals, and their evaluation can be difficult when the number of coefficients increases. It is instead more flexible to simulate from Equation (\ref{eq:TMVN}) and transform back to obtain realisations of \(\boldsymbol{\beta}\). We then compute the empirical covariance matrix and quantiles as estimates of the coefficients' (co)variance and confidence intervals, respectively. This can be performed for any number of constraints up to \(p\), with computational details provided in Supplementary Material C. In this paper, we simulate from Equation (\ref{eq:TMVN}) using the scheme of \citet{Botev2017NormalLawLinear}, which uses exponential tilting in an importance sampling scheme to address acceptance rate issues from previous algorithms \citep{Geweke1991EfficientSimulationMultivariate}. This approach is efficient even for high-dimensional vectors and is implemented in the R software \citep{BotevBelzile2024TruncatedNormalTruncatedMultivariate}.

From the distribution in Equation (\ref{eq:TMVN}), it can be shown that \(\mathbb{E}\left( \mathbf{C}\widehat{\boldsymbol{\beta}} \right) \neq \mathbf{C}\boldsymbol{\beta}^{*}\) \citep{ManjunathWilhelm2021MomentsCalculationDoubly} even when \(\boldsymbol{\beta}^{*}\) would be feasible, \emph{i.e.} that the constrained estimator is biased. On the other hand, the constrained estimator has reduced variance compared to the unconstrained one \citep{BarrSherrill1999MeanVarianceTruncated, Liew1976InequalityConstrainedLeastSquares}, as discussed in Supplementary Material C. This suggests a trade-off between increased bias and reduced variance due to the constraints, which can be advantageous in terms of estimation error. This property is explored in the simulation study below.

%%% DF
\subsection{Degrees of freedom}
Characterising degrees of freedom for a model is essential for various inferential steps, like computing residual variance estimation or model selection criteria, such as the Akaike and Bayesian Information Criteria (AIC and BIC) \citep{BurnhamAnderson2004MultimodelInferenceUnderstanding}. When constraints are applied, degrees of freedom are reduced due to the restrictions imposed on the coefficients. For instance, in the Lasso, degrees of freedom can be shown to be equal to the number of non-zero coefficients \citep{EfronEtAl2004LeastAngleRegression}, while in isotonic regression it corresponds to the number of distinct predictions \citep{MeyerWoodroofe2000DegreesFreedomShaperestricted}. More generally, for any linearly constrained model, the degrees of freedom are
%Equation
\begin{equation} \label{eq:odf}
odf = p - m_a
\end{equation}
where \(p\) is the number of parameters in the model and \(m_{a} \in \{ 0,\ldots,m \}\) is the number of \textit{active} constraints in the fitted model. A constraint represented by the row \(\mathbf{c}_{r}\) (\(r = 1,\ldots,m\)) is active when \(\mathbf{c}_{r}\widehat{\boldsymbol{\beta}} = l_{r}\) or \(\mathbf{c}_{r}\widehat{\boldsymbol{\beta}} = u_{r}\), \emph{i.e.} when an unconstrained model would have yielded an unfeasible solution with respect to the \(r^{th}\) constraint.

We refer to the quantity in Equation (\ref{eq:odf}) as \emph{observed} degrees of freedom (\textit{odf}) because the value of \(m_{a}\) depends on samples \(y_{i}\), and the number of degrees of freedom can be considered a random variable taking values from \(p - m\) to \(p\) \citep{Meyer2013SemiparametricAdditiveConstrained}. Therefore, \emph{odf} might overstate the actual reduction in degrees of freedom induced by the constraints, which can be damaging, in particular for model selection, with a tendency to favour overly complex models.

To better represent the complexity of a model, we should consider the \emph{expected} degrees of freedom (\textit{edf}) similarly to \citet{Meyer2013SemiparametricAdditiveConstrained}:
%Equation
\begin{equation} \label{eq:edf}
edf = p - \sum_{r=0}^{m} r\mathbb{P}\left(m_a = r\right)
\end{equation}
The term \(\sum_{r = 0}^{m}{r\mathbb{P}\left( m_{a} = r \right)}\) represents the expected number of active constraints for the model defined in Equation (\ref{eq:edf}), with the weight \(\mathbb{P}\left( m_{a} = r \right)\) being the probability of having exactly \(r\) active constraints. These weights can be estimated by simulating from a multivariate normal distribution of the unconstrained coefficients and counting the number of constraints that are active in each instance. Here we use the terms \emph{observed} and \emph{expected} degrees of freedom, instead of the generally used \emph{effective} degrees of freedom, to clearly differentiate degrees of freedom computed by Equations (\ref{eq:odf}) and (\ref{eq:edf}).

\section{Practical implementation: the R package 'cirls'}

To make the framework as convenient to use as possible in practice, we provide an R implementation of the framework through the package 'cirls' \citep{MasselotGasparrini2025CirlsConstrainedIteratively}. It provides a \texttt{cirls.fit()} function that embeds the CIRLS algorithm in the standard function \texttt{glm()}. The user can then implement various models using a simple syntax, exploiting various features of the standard R implementation for GLM. This implementation includes the possibility to use several different QP algorithms \citep{GoldfarbIdnani1983NumericallyStableDual, Meyer2013SimpleNewAlgorithm, StellatoEtAl2020OSQPOperatorSplitting}, offering fast and reliable routines with good convergence performance. The package implements inference and model selection through specific methods for standard functions such as \texttt{vcov()}, \texttt{confint()}, and \texttt{logLik()}. This can be combined with other usual functions like \texttt{AIC()} and \texttt{BIC()}.

The package has also been developed for flexibility, providing different interfaces for constraint specification, all complementing each other. Common constraints can easily be provided through a formula interface, while more complex constraint matrices can also be passed directly. These interfaces allow the expansion of the package via user-written interface functions. The software is used in the simulations and practical applications below, as shown in the reproducible code attached to this article.

\section{Simulation study}

In this section, we evaluate the properties of the CIRLS framework introduced in this paper, including the estimation error of the coefficients \(\boldsymbol{\beta}\), the accuracy of the inference, and the definition of expected degrees of freedom defined above.

\subsection{Data-generating mechanisms (DGMs)}

We consider two applications of constrained GLM: i) a non-negative Gaussian regression; and ii) a Poisson regression with five non-decreasing strata. The first DGM generates an association between a Gaussian response and two continuous predictors, with the first coefficient being subject to a non-negative constraint (\(\beta_1 \geq 0\)). The second DGM represents an association between a Poisson response and five strata, with non-decreasing coefficient constraints (\(\beta_j - \beta_{j-1} \geq 0\)). In both DGMs, the generated coefficients are multiplied by a \textit{feasibility} parameter \(\gamma\) that indicates how aligned the true coefficients are with the constraints of the model. When \(\gamma < 0\), the true coefficients are outside the feasible region defined by the constraints, and when \(\gamma > 0\), they are within the feasible region, with \(\gamma = 0\) being exactly at the boundary. This allows evaluating the performance of CIRLS under both appropriate and wrong constraints. For both DGMs, we draw \(n_{sim}=10\,000\) samples of size \(n = 100,500\) and low signal-to-noise ratios to emulate a low-power situation. Both data-generating mechanisms are summarised in Table \ref{tab:DGM}. 

%%%%% TABLE

\begin{table}
\centering
\caption{Summary of the two data-generating mechanisms (DGM) for the simulation study. For both DGMs, we perform 10 000 simulations of datasets with sample size \(n = 100,500\).}
\resizebox{\textwidth}{!}{
\begin{tabular}{lcc}
\hline
& \textbf{Non-negative regression} & \textbf{Non-decreasing strata} \\
\hline

\(p\) & 2 & 5 \\

$\mathbf{x}_i$ & MVN$(0,\boldsymbol{\Sigma})$ & $\mathrm{Unif}\{1,\ldots,5\}$ \\

\(\boldsymbol{\beta}\) & \([\gamma; 1]\) & \(\beta_j = \gamma\left(1 + \exp(-50 j)\right)\) \\

$y_i$ & $N(5 + \mathbf{x}_i \boldsymbol{\beta}, 50)$ & $\mathrm{Poisson}(\exp(\mathbf{x}_i \boldsymbol{\beta}))$ \\

CIRLS constraints & $\beta_1 \ge 0$ & $\beta_{j+1} - \beta_j \ge 0$ \\
\hline
\end{tabular}
}
\label{tab:DGM}
\end{table}

% After the table
Estimation performance of the CIRLS fit is evaluated by computing the absolute bias (\(|\text{Bias}|\)), standard error (SE) and root mean squared error (RMSE). These three performance measures are computed on a constrained and an unconstrained fit, and then we report their differences (\textit{e.g.} \(\Delta \text{RMSE} = \text{RMSE} - \text{RMSE}^*\) where \(\text{RMSE}^*\) is the unconstrained one). Inference from the TMVN is evaluated by computing the relative error of coefficient variance as well as 95\% confidence interval coverage, both estimated through 10 000 simulations by the exponential tilting algorithm of \citet{Botev2017NormalLawLinear}. Finally, the degrees of freedom formulae (Equations \ref{eq:odf} and \ref{eq:edf}) are compared to the true degrees of freedom computed from the simulations \citep{ShenEtAl2004AdaptiveModelSelection, Ye1998MeasuringCorrectingEffects}. The specific formulae for all these performance criteria can be found in Supplementary Material D (Equations D.1, D.2, and D.3).

\subsection{Results}

This section shows selected results of the simulation study for \(n = 500\). Results for \(n=100\) are provided in Supplementary Material D, confirming the conclusions drawn for this section. Supplementary Material D also reports additional measures of performance.

\subsubsection{Estimation error}

\begin{figure}
  \includegraphics[width=1\textwidth]{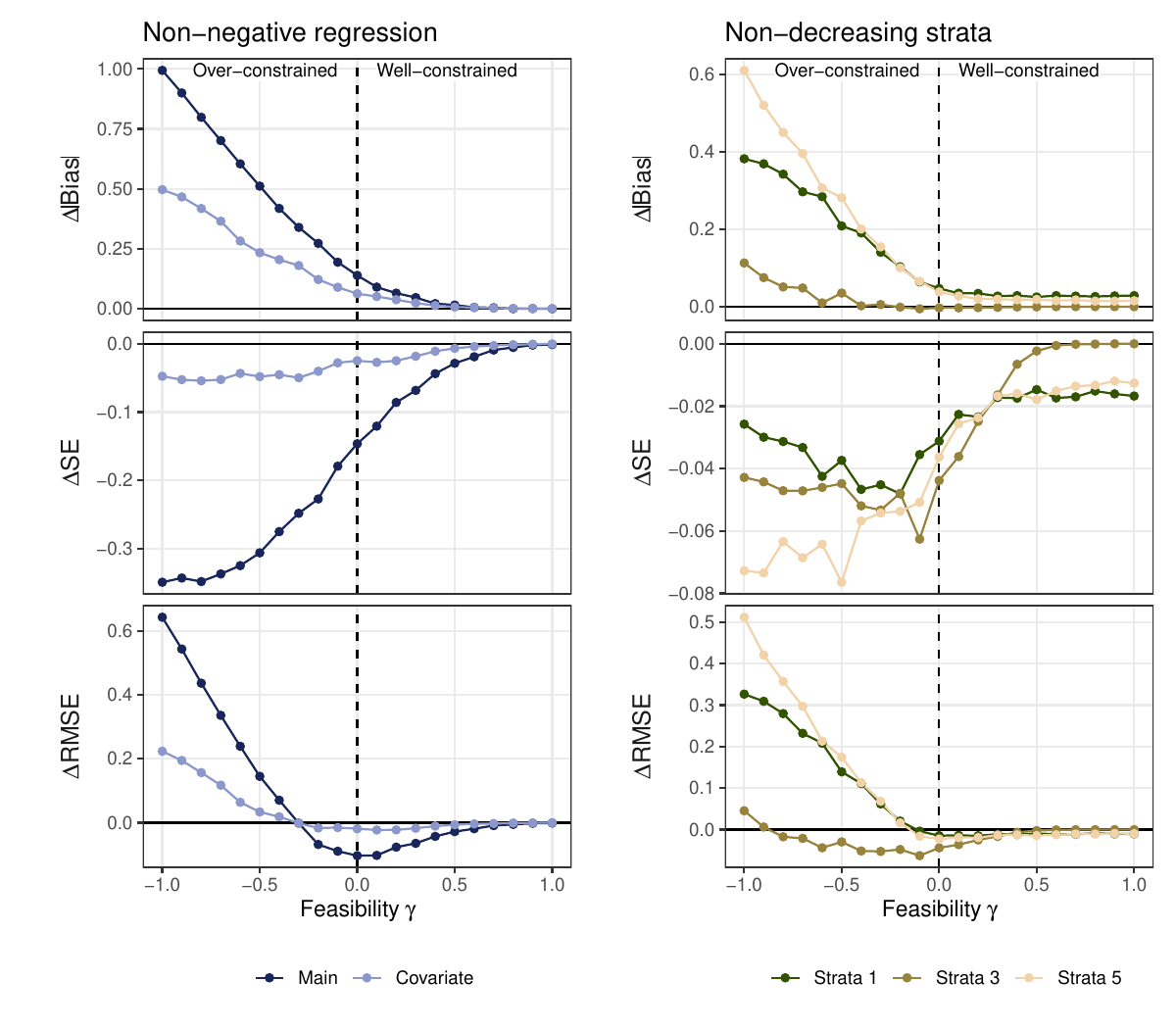}
  \caption{Difference in absolute bias (top row), standard error (middle row) and root mean squared error (RMSE, bottom row) between CIRLS and an unconstrained GLM. Positive (negative) \(\Delta\) indicates worse (better) performance of CIRLS compared to the unconstrained estimate. The x-axis corresponds to the feasibility parameter \(\gamma\), with negative values indicating an over-constrained model and positive values an appropriately constrained model. In the first DGM (left), the ‘main’ coefficient is \(\beta_1\) and ‘covariate’ is \(\beta_2\), while in the second (right), strata 1, 3, and 5 respectively correspond to \(\beta_1\), \(\beta_3\) and \(\beta_5\) in \ref{tab:DGM}.}
  \label{fig:simRMSE}
\end{figure}

Figure \ref{fig:simRMSE} shows \(\Delta|\text{Bias}|\), \(\Delta\text{SE}\) and \(\Delta \text{RMSE}\) versus the feasibility parameter \(\gamma\) for \(n=500\). In both DGM and for all coefficients, there is a clear decreasing trend in \(\Delta|\text{Bias}|\) and an increase in \(\Delta\text{SE}\). They both converge towards zero when \(\gamma\) increases above 0, indicating that the CIRLS provide similar results to an unconstrained model when constraints are not restrictive. This results in inverse-J-shaped curves of \(\Delta \text{RMSE}\), where the RMSE is high for widely over-constrained models, then decreases with the bias, and finally becomes negative closer to the boundary (\(\gamma=0\)), indicating an improvement in RMSE for CIRLS compared to an unconstrained model. This pattern is observed for all coefficients in Figure \ref{fig:simRMSE}, but with various amplitudes. In the first DGM, \(\Delta \text{RMSE}\) also displays this pattern for the unconstrained covariate but with a lower amplitude than the main coefficient. In the second DGM, strata 1 and 5 are more sensitive to the constraints being well defined, with the bias and RMSE increasing rapidly when \(\gamma\) decreases.

Importantly, the improvement in RMSE for CIRLS also happens for several negative feasibility parameters, \textit{i.e.} where the model is actually slightly over-constrained. In the first DGM, for example, \(\Delta \text{RMSE}\) is negative even when the true coefficient \(\beta_1=-0.3\), while it is negative in almost all situations for Strata 3 in the second DGM. The lowest value of RMSE is generally close to the boundary of the feasible region, which outlines the importance of carefully choosing the constraints. In summary, appropriately chosen constraints, even when slightly over-constraining, can be beneficial for the estimation performance of a GLM.

\subsubsection{Uncertainty assessment}

\begin{figure}
  \includegraphics[width=1\textwidth]{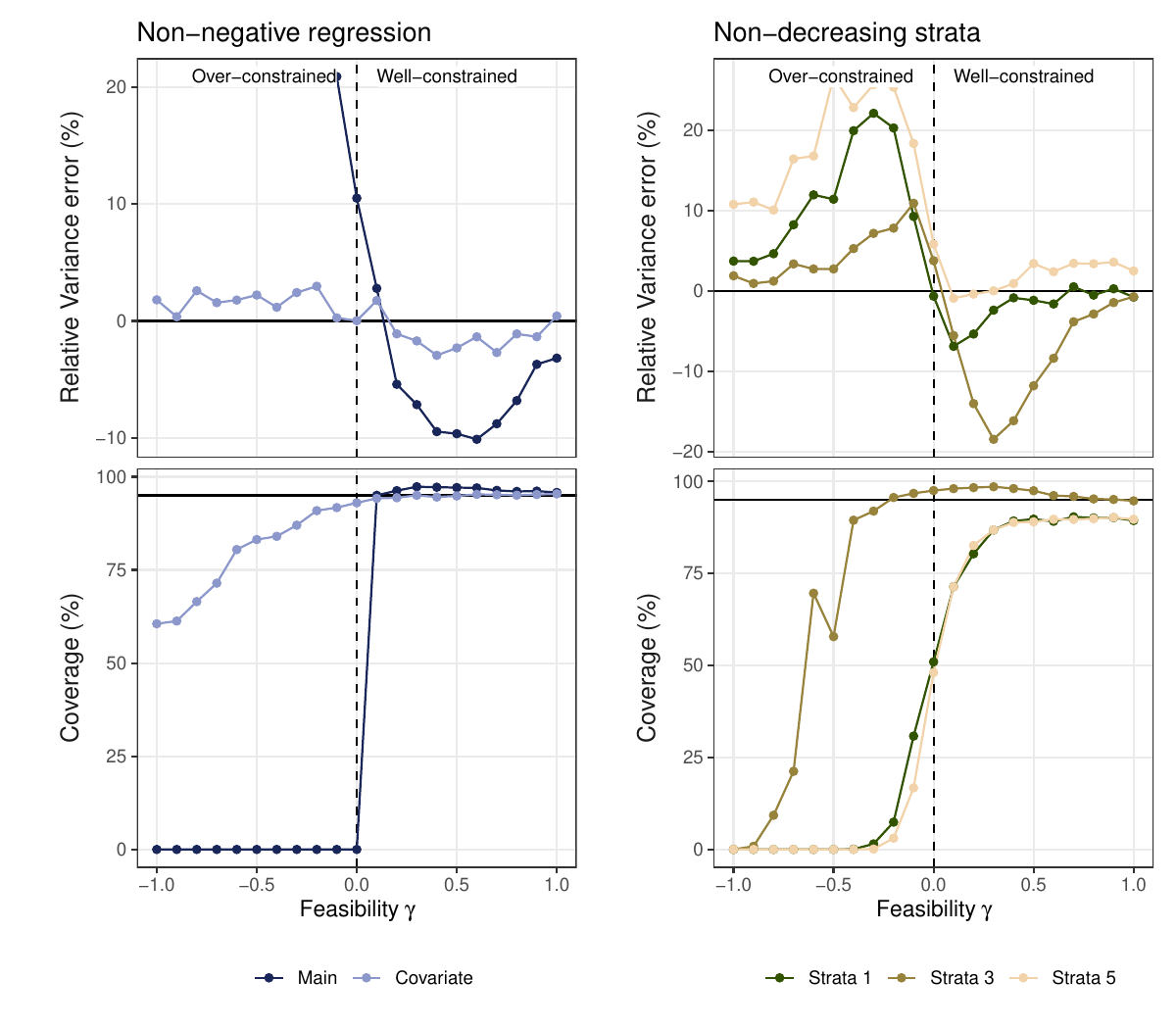}
  \caption{Evaluation of the Truncated Multivariate Normal (TMVN) inference for constrained GLMs versus the feasibility parameter \(\gamma\). The top row shows the Relative Variance error, i.e. the ratio between the average coefficient standard error and the standard deviation of estimated coefficients across simulations. The bottom row shows the coverage of a 95\% confidence interval.}
  \label{fig:simCoverage}
\end{figure}

Figure \ref{fig:simCoverage} shows the error in the inference procedures according to the feasibility parameter \(\gamma\). In both DGMs, the variance is overestimated when the model is over-constrained, and slightly underestimated for appropriately constrained models, with the amplitude of the error depending on how constrained the coefficient is. However, except for the main coefficient of the first DGM when \(\gamma\) is low, the error remains within 30\%. The error is often negligible when the true linear predictor is near the boundary of the feasible region (\(\gamma=0\)). 

Similarly, the coverage is null when the model is over-constrained since, by definition, it cannot include the true coefficient value. In the first DGM (bottom left panel), for the lowest feasibility parameter, this also brings the coverage of the covariate coefficient down due to the added bias (Figure \ref{fig:simRMSE}). Note that, when correcting for the bias, the coverage is close to the nominal level even for some over-constrained models, indicating an acceptable confidence interval width (see Supplementary Material D). However, when appropriately constrained, the coefficient immediately reaches 95\% coverage with a slight over-coverage for the feasibility parameter between \(\gamma=0.2\) and \(\gamma=0.6\). In the second DGM (bottom right panel), the same pattern is visible, although smoother since the constraints are not directly on the coefficient value but on their relation. Note that the confidence intervals of Strata 1 and 5 still display some under-coverage even for the highest values of \(\gamma\). This is because these strata are involved in strongly active constraints by having coefficients defined almost exactly on the boundary of the constraint (see Figure D.1 in Supplementary Material D). However, with the TMVN approach, the constraint boundaries have null probability leading to under-coverage.

\subsubsection{Degrees of freedom}

\begin{figure}
  \includegraphics[width=1\textwidth]{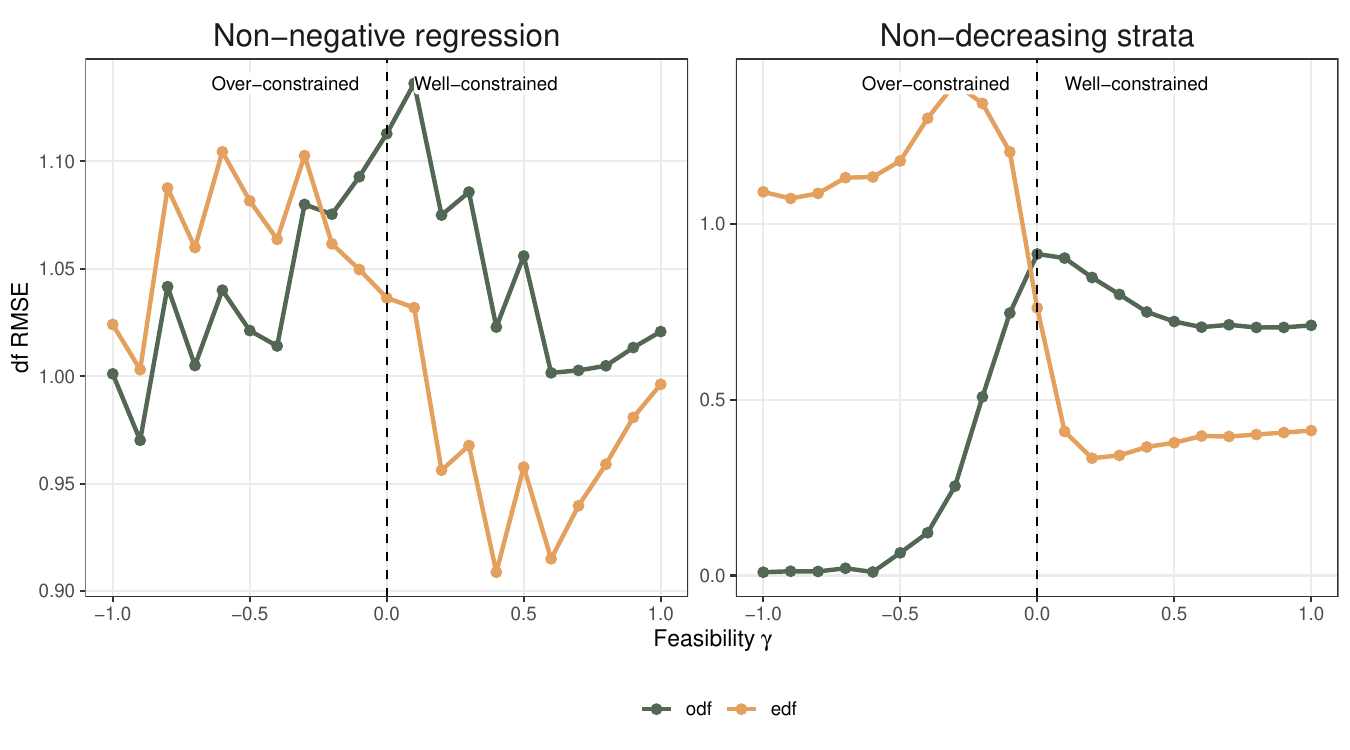}
  \caption{Root mean squared error (RMSE) of observed degrees of freedom (odf) and expected degrees of freedom (edf) in estimating the actual degrees of freedom of the constrained GLM depending on the feasibility parameter \(\gamma\). The exact definition of actual degrees of freedom is provided in Supplementary Material.}
  \label{fig:simdf}
\end{figure}

Figure \ref{fig:simdf} shows the RMSE between both \textit{odf} and \textit{edf}, and the actual degrees of freedom calculated from the simulated observations. It indicates that \textit{edf} tends to be a better estimate of model complexity than \textit{odf} for well-constrained models, as well as slightly over-constrained ones. The difference is much more important in the second DGM than in the first one, because several constraints are involved. This suggests that the usefulness of \textit{edf} might increase with more complex constraints. Note that \textit{odf} is generally unbiased but unstable since it ignores the randomness in the number of active constraints. Instead, \textit{edf} accounts for randomness and has much lower standard error, especially close to the boundary between feasible and unfeasible models, but at the cost of some bias, especially for widely over-constrained models (see Supplementary Material C).

%%%%%% Applications
\section{Real-world case study: temperature-related mortality}

Heat and cold are important environmental health risks that are subject to intensive epidemiological investigation \citep{GasparriniEtAl2015MortalityRiskAttributable,MasselotEtAl2023ExcessMortalityAttributed}. Methods for estimating how mortality varies with temperatures are complex \citep{MasselotGasparrini2025ModellingExtensionsMultilocation} and often result in artefacts that may make these associations difficult to interpret. It is now well established that the risk of mortality increases at both ends of the temperature distribution \citep{AlahmadEtAl2025AreThereMore}. Therefore, it is appropriate to add a convexity constraint on this association.

We illustrate how these convexity assumptions can be introduced using a real dataset of daily time series of temperature and mortality in Florence, Italy for the period 2011-2021. Temperature was extracted from the global ERA5-Land dataset, while mortality is extracted from the public records from ISTAT. Following established methods, we consider a typical Poisson model for daily deaths $y_i$ with a canonical log link function as in \citet{BhaskaranEtAl2013TimeSeriesRegression}:

\begin{equation}
    \log(\mathbb{E}(y_i)) = \alpha + s_1(x_i, \boldsymbol{\beta}) + s_2(t_i, \boldsymbol{\theta}) + \mathbf{z}_i^T \boldsymbol{\zeta}
\end{equation}

where \(s_1(x_i, \boldsymbol{\beta})\) is a nonlinear term of a four-day moving average of temperature parametrised by the vector \(\boldsymbol{\beta}\). The second term \(s_2(t_i, \boldsymbol{\theta})\) is a natural cubic spline of time \(t_i\) with seven degrees of freedom per year to capture the trend and seasonality of daily deaths, while \(\mathbf{z}_i\) is a vector of dummy variables for the day of the week.

We consider two different models for \(s_1(x_i, \boldsymbol{\beta})\): 1) a binning model where the values of \(x_i\) are discretised into \(2^\circ\)C wide bins, and b) a smooth model where \(s_1(x_i, \boldsymbol{\beta})\) is represented by quadratic B-splines with 4 internal knots placed at the quintiles. In both cases, convexity of \(s_1(x_i, \boldsymbol{\beta})\) is imposed by a convexity constraint directly on the \(\boldsymbol{\beta}\) coefficients, \textit{i.e.} \(\beta_j - 2\beta_{j-1} + \beta_{j-1} \geq 0\) \citep{PyaWood2015ShapeConstrainedAdditive}. The full details are shown in Supplementary Material E.

\begin{figure}
  \includegraphics[width=1\textwidth]{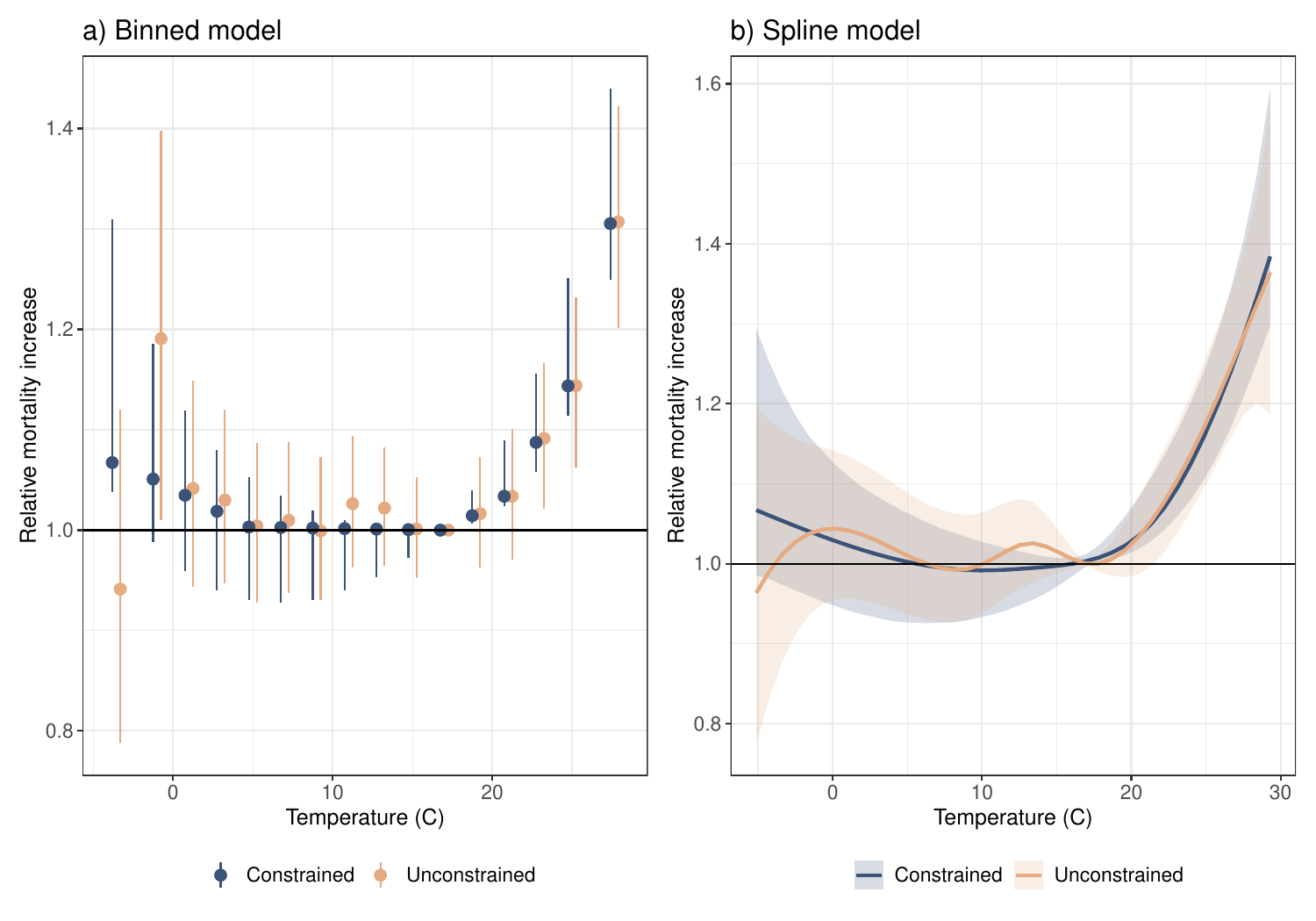}
  \caption{Estimated associations between daily mortality and temperature in Florence from a Poisson model with a) a binning model and b) a quadratic spline model. The constrained association is estimated with CIRLS with a convexity constraint on the coefficients of temperature. Vertical bars and shaded bands represent 95\% confidence intervals.}
  \label{fig:firenze}
\end{figure} 

Figure \ref{fig:firenze} shows unstable unconstrained estimate with an implausible decrease of the risk for extreme cold temperatures as well as an unclear minimum mortality temperature (an important epidemiological summary). In contrast, constrained models result in a more plausible curve with a wide comfort zone of no increase in risk, and a smoother risk increase on the cold part. Also note that where the unconstrained model agrees with the convexity assumption (for temperatures above \(20^\circ\)C), the two estimates are very close to each other, but the CIRLS estimate has narrower confidence intervals.

%%%% Discussion
\section{Discussion}

In this contribution, we developed a constrained iteratively reweighted least squares (CIRLS) framework to estimate GLMs with constrained coefficients. The development includes inferential procedures, with the distribution of constrained coefficients for inference and the representation of the model’s complexity through its degrees of freedom. The whole framework is implemented in an easy-to-use R package to allow widespread application of the framework. A simulation study suggests that appropriately defined constraints improve the accuracy of coefficient estimation through decreased estimation variance, despite the introduction of a slight bias. Importantly, this property was also observed when constraints excluded the true generated coefficients to a certain extent. A case study applies CIRLS for constraining shapes of exposure-response relationships, illustrating the flexibility of the method.

Experiments with the CIRLS algorithm for this contribution have shown that the CIRLS algorithm converges quickly. CIRLS can be shown to be a Sequential Quadratic Program (SQP), a general class of algorithms that can solve nonlinear constrained optimisation problems rather efficiently \citep{BoggsTolle1995SequentialQuadraticProgramming}. Some work has shown that such algorithms converge quadratically to a local optimum \citep{NocedalWright2006NumericalOptimization} under some conditions that are broadly met in the constrained GLM context. In CIRLS, each step of the algorithm uses highly efficient algorithms to solve quadratic programs \citep{GoldfarbIdnani1983NumericallyStableDual, StellatoEtAl2020OSQPOperatorSplitting}, ensuring a low computational burden. We refer the reader to Supplementary Material B, which develops this point further. 

We have proposed a framework for the inference of coefficients that draws from TMVN distributions, which showed acceptable properties in our simulations. However, this framework presents important limitations which can hamper inference in some practical applications. First, our simulations show that it leads to some under-coverage when true coefficients are close to the constraints. Second, the framework limits inference to applications where the number of constraints \(m\) is no larger than the number of variables \(p\), otherwise the back-transformation cannot be computed. Although this allows many practical cases, this prevents inference in some useful applications, such as S-shaped regression \citep{Meyer1999ExtensionMixedPrimal} or the Lasso \citep{Tibshirani1996RegressionShrinkageSelection}. In the \(m>p\) case, it is possible to combine the proposed inference with an accept-reject algorithm in which the TMVN sampling is performed with a selection of \(p\) constraints, and retains only the samples that satisfy the remaining \(m-p\) constraints \citep{Geweke1996BayesianInferenceLinear}. However, the practicality of this approach depends on the acceptance rate, which can be prohibitively low when \(m-p\) is high.

Another limitation of the inference approach is that it necessitates fitting the equivalent unconstrained model used in Equation \ref{eq:TMVN}. This typically precludes inference in \(p\geq n\) applications in which the constrained GLM can be fitted but the unconstrained cannot, which includes the Lasso as well as isotonic regression. Previous works have proposed inference for individual coefficients for the Lasso and variable selection more generally \citep{LeeEtAl2016ExactPostselectionInference, TaylorTibshirani2015StatisticalLearningSelective}. It takes advantage of the result that the distribution of individual coefficients, conditional on other coefficients, is a truncated univariate normal \citep{Horrace2005ResultsMultivariateTruncated}. Although this requires more complex formulas than what is proposed in the present paper, these results can represent a way forward to generalise the inference from CIRLS fits.

The Bootstrap can represent a flexible alternative for inference in the more extreme applications of CIRLS as described above. It has been used extensively for the Lasso \citep{HastieEtAl2015StatisticalLearningSparsity}, other constrained regression models \citep{MasselotEtAl2023ConstrainedGroupwiseAdditive}, as well as for the estimation of the mixture probabilities in degrees of freedom computation \citep{Meyer2013SemiparametricAdditiveConstrained}. However, it is more computationally demanding, and its application is less trivial in constrained estimation. Parametric Bootstrap, as used in other applications \citep{Meyer2003TestLinearConvex}, can only be performed when the unconstrained model can be fitted. Additionally, the Bootstrap generally tends to be inconsistent for inference when parameters are close to the boundary of their feasible space \citep{Andrews2000InconsistencyBootstrapWhen} and therefore requires more complex procedures \citep{Li2025ProximalBootstrapConstrained}.

Future work must focus on the extension of inference to hypothesis testing. In many applications, there is interest in testing how binding the constraints are and whether they represent good assumptions on the relationship. Examples include testing the positivity of coefficients \citep{Davis1978BayesianAnalysisLinear} and the shape of a nonlinear association \citep{Meyer2003TestLinearConvex, SenMeyer2017TestingLinearRegression}. Additionally, it is of interest to provide significance tests for coefficients in the presence of constraints. Finally, future work can expand the applicability of CIRLS, allowing nonlinear constraints, including, for instance, ridge-type constraints useful for smoothing coefficients. Indeed, we have shown the connection between CIRLS and SQP algorithms, which can be exploited to include more general constraints.

In conclusion, the CIRLS algorithm represents a simple and flexible framework for GLMs with linear constraints on the coefficients. Such constraints can help the analysis of biomedical or epidemiological data in situations in which a classical GLM would be difficult to fit, with clear improvements on the estimation of coefficients when the constraints are appropriately set. The proposed inference and degrees of freedom allow the analyst to perform the usual tasks of model selection and confidence interval computation.

%%% Acknowledgements (if any)
%%% ------------------------------------------
\section*{Acknowledgements}
The methods proposed in this paper are implemented in the package `cirls` for the open-source software R. The simulation study and illustrative applications can be fully replicated, with R code and data available on GitHub (https://doi.org/10.5281/zenodo.18746187).

%%% Funding (if any)
%%% ------------------------------------------
\section*{Funding}
This research is supported by the Medical Research Council (MRC) of the United Kingdom (grant MR/X029476/1).

%%% Appendix (if any)
%%% ------------------------------------------
%%%%%%%%%%%%%%%%%%%%%%%%%%%%%%
% Appendix

\appendix

%%%%%%%%%% Section: Model

\section{Generalised linear models with linear constraints}

\subsection*{Model definition}

Following standard generalised linear models (GLM) theory \citep{McCullaghNelder1989GeneralizedLinearModels,Wood2017GeneralizedAdditiveModels}, we assume a \(n\)-dimensional response vector \(\mathbf{y}\) with elements \(y_i\) following a distribution belonging to the exponential family, that is:
\begin{equation} \label{eq:expfamily}
    p_\theta(y_i)= \exp\left[ \left(y_i\theta_i - b(\theta_i) \right) / a(\phi) + c(y_i,\phi) \right]
\end{equation}
where the functions \(a(.)\), \(b(.)\) and \(c(.)\) depend on the specific distribution and \(\phi\) is the dispersion parameter. The values of these functions and parameters for a list of common distributions is provided by \citet{McCullaghNelder1989GeneralizedLinearModels}. 

Derivation of \ref{eq:expfamily} shows that \(\mathbb{E}(y_i) = b'(\theta_i)\) that is the first derivative of \(b(\theta_i)\). Naming \(\mu_i=\mathbb{E}(y_i)\), we link it to the \(p\)-dimensional predictor vector \(\mathbf{x}_i\) through \(\theta_i\). In this contribution, we look to estimate the model with \(m\) constraints, i.e. the model:
\begin{equation} \label{eq:cGLM2}
\begin{aligned}
    g(\mu_i) = \mathbf{x}_i^{T}\boldsymbol{\beta} \\
    \text{subject to} \quad & \mathbf{l} \le \mathbf{C}\boldsymbol{\beta} \le \mathbf{u}
\end{aligned}
\end{equation}
where \(\boldsymbol{\beta}\) is a \(p\)-dimensional vector of regression coefficients and \(g()\) a link function. The latter can often be chosen such that \(g(b'(\theta_i))=\theta_i\). Constraints are represented by a \(m \times p\) constraint matrix \(\mathbf{C}\) where \(m\) is the number of constraints, with bound vectors \(\mathbf{l}\) and \(\mathbf{u}\) (both of dimension \(m\)). The constraints include equality constraints with \(l_r=u_r\) for any constraint \(r\), as well as one-side constraints with \(l_r=-\infty\) or \(u_r = \infty\).

\subsection*{Examples of constraints}

Here, we provide a few examples of linear constraints that can be applied with the constrained GLM.

\subsubsection*{Non-negative least squares}
Non-negative least squares can be extended to GLM in this framework, with the constraints that \(\beta_j \geq0\) with \(j=1,...,p\) \citep{McDonaldDiamond1990FittingGeneralizedLinear}. In matrix form, we have \(\mathbf{C} = \mathbf{I}_p\) (the identity matrix), \(\mathbf{l} = \mathbf{0}\) and \(\mathbf{u}=\infty\).

\subsubsection*{Isotonic regression}
Isotonic regression is a non-parametric regression that seeks to estimate a non-decreasing association between a response and a predictor \citep{BarlowBrunk1972IsotonicRegressionProblem}. In isotonic regression, there is typically one parameter for each data point (\(p=n\)) which is fitted with the constraints \(\beta_j \geq \beta_{j-1}\). We therefore have: 
\begin{equation}\label{eq:diffmat}
    \mathbf{C} = \begin{pmatrix}
        -1 & 1 & 0 & \cdots & 0 & 0\\
        0 & -1 & 1 & \cdots & 0 & 0\\
        \vdots & \vdots & \vdots & \ddots & \vdots & \vdots\\
        0 & 0 & 0 & \cdots & -1 & 1 
    \end{pmatrix}
\end{equation}
with \(\mathbf{l} = \mathbf{0}\) and \(\mathbf{u}=+\infty\). Here \(\mathbf{C}\) is of dimension \(p-1\times p\). The opposite constraint (non-increasing) is set up identically, by either swapping the values of \(\mathbf{l}\) and \(\mathbf{u}\), or swapping the -1 and 1 in each row of \(\mathbf{C}\).

\subsubsection*{Convex regression}
This extends isotonic regression by forcing a convex association. Convexity of the coefficients \(\beta_j\) is enforced with the \(p-2\times p\) constraint matrix:
\begin{equation} \label{eq:cvxmat}
    \mathbf{C} = \begin{pmatrix}
        1 & -2 & 1 & \cdots & 0 & 0\\
        0 & 1 & -2 & \cdots & 0 & 0\\
        \vdots & \vdots & \vdots & \ddots & \vdots & \vdots\\
        0 & 0 & 0 & \cdots & -2 & 1 
    \end{pmatrix}
\end{equation}
with \(\mathbf{l} = \mathbf{0}\) and \(\mathbf{u}=\infty\). Similarly, concave coefficients are obtained by swapping the values of \(\mathbf{l}\) and \(\mathbf{u}\).

\subsubsection*{Constrained splines}
Linear constraints can be used to impose shapes on smooths that are based on B-splines. In that context, the linear predictor is \(\mathbf{x}_i^{(q)}\) that contains B-splines bases of order \(q\) indexed by \(L\) knots \(\xi_l\). The \(k^{th}\) derivative of a function derived from B-spline bases can be shown to be \citep{ZhouWolfe2000DerivativeEstimationSpline}:
\begin{equation}
    s^{(k)}(x_i)= {\mathbf{x}_i ^{(q-k)}}^T \mathbf{D}^{(k)} \boldsymbol{\beta} 
\end{equation}
where \(\mathbf{x}_i ^{(q-k)}\) is the vector containing B-splines bases of order \(q-k\) evaluated at \(x_i\) and the difference matrix is the product of \(k\) matrices \(\mathbf{D}^{(k)}=\mathbf{E}_k \ldots \mathbf{E}_1 \), with:
\begin{equation}
    \mathbf{E}_k = \begin{pmatrix}
        \frac{1}{\xi_1 - \xi_{1-q + k}} & \frac{-1}{\xi_2 - \xi_{2-q + k}} & 0 & \cdots & 0 \\
        0 & \frac{1}{\xi_2 - \xi_{2-q + k}} & \frac{-1}{\xi_3 - \xi_{3-q + k}} & \cdots & 0  \\
        0 & 0 & \frac{1}{\xi_3 - \xi_{3-q + k}} & \cdots & 0  \\
        \vdots & \vdots & \vdots & \ddots & \vdots \\
        0 & 0 & 0 & \cdots & \frac{1}{\xi_3 - \xi_{L-q + k}} \\
    \end{pmatrix}
\end{equation}
Note that the \(\mathbf{E}_k\) matrices are simply weighted versions of (\ref{eq:diffmat}) depending on the knot placement. Similarly, \(\mathbf{D}^{(2)}\) will be a weighted version of (\ref{eq:cvxmat}). 

The expressions above can be used to impose monotonicity and convexity constraints by constraining the first and second derivative, respectively, to be non-negative. Since B-spline bases in \(\mathbf{x}_i ^{(q-k)}\) are always non-negative by construction, this means that such constraint can be enforced by \(\mathbf{C} = \mathbf{D}^{(1)}\) for monotonicity and \(\mathbf{C} = \mathbf{D}^{(2)}\)
for convexity or concavity. Note that when the knots are all equidistants, the constraint matrix reduces to  (\ref{eq:diffmat}) or  (\ref{eq:cvxmat}). The bound vectors \(\mathbf{l}\) and \(\mathbf{u}\) are identical to isotonic and convex regressions above.

\subsubsection*{Compositional regression}
Compositional data \citep{Aitchison1982StatisticalAnalysisCompositional} refer to a set of variables that represent the relative parts of a quantity, \textit{i.e.} a set of \(p\) variables such that \(\sum_{j=1}^p z_{ij} = 1\) (the sum can be any constant value).  Such variables can be used in a GLM by using \(x_{ij}=\log(z_{ij})\) as predictors, and setting the constraint that \(\sum_{j=1}^p \beta_j = 0\) \citep{AitchisonBacon-Shone1984LogContrastModels}. In matrix terms, this means that we have \(\mathbf{C} = \mathbf{1}^T\), where \(\mathbf{1}^T\) is a \(p\)-dimensional vector of ones, and \(\mathbf{l} = \mathbf{u} = 0\). This constraint can be extended to several compositions in which case \(\mathbf{C}\) becomes block-diagonal, with each block a vector of ones \citep{ShiEtAl2016RegressionAnalysisMicrobiome}.

\subsubsection*{Lasso}
The least absolute shrinkage and selection operator (Lasso) is a popular technique to deal with high-dimensional predictors \citep{Tibshirani1996RegressionShrinkageSelection}. The idea of the Lasso is to constraint the sum of absolute coefficients, which forces a number of coefficient to have zero value. The Lasso constraint is 
\begin{equation} \label{eq:Lasso}
    \sum_j |\beta_j| \leq t
\end{equation}
with \(t\) a scalar that determines the stringency of the constraint and is typically chosen by cross-validation (usually through the Lagrange multiplier \(\lambda\) of the constraint). The constraint (\ref{eq:Lasso}) can actually be expressed as a set of \(2^p\) linear constraints representing all possible signs of \(\beta_j\) coefficients. For \(p=2\) (say), we therefore have
\begin{equation*}
    \mathbf{C} = \begin{pmatrix}
        1 & 1 \\
        -1 & 1 \\
        1 & -1 \\
        -1& -1
    \end{pmatrix}
\end{equation*}
with \(\mathbf{l} = -\infty\) and \(\mathbf{u}\) is a 4-dimensional vector containing the value \(t\).
%%%%%%%%%% Section: details on the algorithm
\section{The CIRLS algorithm}

Here, we provide the full details about the CIRLS algorithm and attempt to provide reassurance regarding its convergence by linking it to the more general class of sequential quadratic programming (SQP) algorithms.

%%% This is where we present the full algorithm
\subsection*{Constrained Iteratively-Reweighted Least-Squares (CIRLS)}

The CIRLS algorithm seeks to estimate \(\boldsymbol{\beta}\) under constraints as expressed in Equation \ref{eq:cGLM2}. It is an extension to the (unconstrained) iteratively-reweighted least-squares (IRLS) and is fully laid out in Algorithm 1. Notice that it uses Fisher scoring weights instead of a full Newton descent step. We rewrite the objective function in Equation (\ref{eq:CIRLS2}) to match a typical Quadratic Programming (QP) format \citep{BoydVandenberghe2004ConvexOptimization}. 

As in the unconstrained IRLS algorithm, we consider it has converged when the relative decrease in deviance \(D(\boldsymbol{\mu}^{[k]})=2\left(l(\hat{\boldsymbol{\beta}}_{max}) - l(\hat{\boldsymbol{\beta}}) \right)\) is small. Here, \(l(\hat{\boldsymbol{\beta}})\) is the log-likelihood of the model, with \(l(\hat{\boldsymbol{\beta}}_{max})\) the saturated likelihood with one parameter per data point. See \citet{McCullaghNelder1989GeneralizedLinearModels} for details on the deviance.

%%%%% The algorithm
\begin{algorithm}[h] \label{algo:CIRLS}
    \caption{The CIRLS algorithm}
    \begin{enumerate}
        \item Initialise \(\boldsymbol{\mu}^{[0]} = \mathbf{y} + \boldsymbol{\delta}\) and \(\boldsymbol{\eta}^{[0]} = g(\boldsymbol{\mu}^{[0]})\). \(\boldsymbol{\eta}^{[0]}\) is an initial estimate of the linear predictor and \(\boldsymbol{\delta}\) is a vector of small constants to ensure \(\boldsymbol{\eta}^{[0]}\) is defined.
        \item Iterate the following steps over \(k\)
        \begin{enumerate}
            \item Compute the pseudo-response vector \(\mathbf{z} = \mathbf{G}(\mathbf{y} - \boldsymbol{\mu}) + \boldsymbol{\eta}^{[k]}\) (\(\mathbf{G}\) is a diagonal matrix with elements \(G_{ii} = g'(\mu_i^{[k]})\)), and pseudo-weight diagonal matrix \(\mathbf{W}\) with elements \(W_{ii} = (g'(\mu_i^{[k]})^2V(\mu_i^{[k]}))^{-1}\) where \(V(\mu_i^{[k]})\) is the variance component of the distribution.
            \item Estimate \(\hat{\boldsymbol{\beta}}\) by solving the following quadratic program (QP)
                \begin{equation} \label{eq:CIRLS2}
                \begin{split}
                \hat{\boldsymbol{\beta}} & = \min_{\boldsymbol{\beta}} \left( \mathbf{z} - \mathbf{X}\boldsymbol{\beta} \right)^{T} \mathbf{W} \left( \mathbf{z} - \mathbf{X}\boldsymbol{\beta} \right) \\
                    & = \min_{\boldsymbol{\beta}}
                       \left(
                           \boldsymbol{\beta}^T \mathbf{X}^T \mathbf{W} \mathbf{X} \boldsymbol{\beta}
                           - 2 \mathbf{z}^T \mathbf{W} \mathbf{X} \boldsymbol{\beta}
                           + \mathbf{z}^T \mathbf{z}
                       \right) \\
                       \text{s.t.} & \quad \mathbf{l} \le \mathbf{C}\boldsymbol{\beta} \le \mathbf{u}
                \end{split}
                \end{equation}
            \item Update \(\boldsymbol{\eta}^{[k+1]} = \mathbf{X} \hat{\boldsymbol{\beta}}\) and \(\boldsymbol{\mu}^{[k+1]} = g^{-1}(\boldsymbol{\eta}^{[k+1]})\)
        \end{enumerate}
        \item Stop when the change in deviance is below a predefined threshold, that is \(\left(D(\boldsymbol{\mu}^{[k]}) - D(\boldsymbol{\mu}^{[k-1]}\right) / D(\boldsymbol{\mu}^{[k]}) \leq \epsilon\), with \(\epsilon\) a small constant.
            
    \end{enumerate}
\end{algorithm}

%%%%%%%%%%%%%%%% Subsection SQP
\subsection*{Sequential Quadratic Programming}

CIRLS belongs to the class of Sequential Quadratic Programming (SQP) algorithms, which optimise general constrained nonlinear problems of the form:
%%% Equation: general optimisation
\begin{equation} \label{eq:optim}
  \begin{split}
    \min_{\boldsymbol{\beta}} & \quad f(\boldsymbol{\beta}) \\
    \text{s.t.} & \quad c_r(\boldsymbol{\beta}) \geq 0 
  \end{split}
\end{equation}
where \(f(.)\) is an objective function and \(c_r(.)\) (\(r=1,...,m\)) are constraint functions. The problem in equation (\ref{eq:optim}) is written generally to keep notation simple, but it encompasses equality constraints and bounds different than zero.

To solve Equation (\ref{eq:optim}), an SQP starts with an initial value \(\boldsymbol{\beta}^{[0]}\), and iteratively update it through QPs. More specifically, the update \(\mathbf{b}\) is the solution of the QP generally written as \citep{ NocedalWright2006NumericalOptimization}:
%%% Equation: QP
\begin{equation} \label{eq:SQP}
  \begin{split}
    \min_{\mathbf{b}}  & \quad  f(\boldsymbol{\beta}^{[k]}) + \nabla f(\boldsymbol{\beta}^{[k]})^T \mathbf{b} + \mathbf{b}^T \nabla^2 L(\boldsymbol{\beta}^{[k]}) \mathbf{b} \\
    \text{s.t.}  & \quad  \nabla c_r(\boldsymbol{\beta}^{[k]}) \mathbf{b}^T + c_r(\boldsymbol{\beta}^{[k]}) \geq 0 
  \end{split}
\end{equation}
where \(\nabla\) indicates the gradient vector and \(\nabla^2\) the Hessian evaluated at \(\boldsymbol{\beta}^{[k]}\). The term \(L(\boldsymbol{\beta}^{[k]})\) refers to the Lagrangian of the optimisation problem (\ref{eq:optim}), \textit{i.e.} \( L(\boldsymbol{\beta}^{[k]}) = f(\boldsymbol{\beta}^{[k]}) - \sum_r \lambda_r  c_r(\boldsymbol{\beta}^{[k]}) \) where the \(\lambda_r (r=1,...,m)\) are the Lagrange multipliers. However, note that when the constraints are linear - which is our case - the Lagrangian \(\nabla^2 L(\boldsymbol{\beta}^{[k]})\) simply reduces to \(\nabla^2 f(\boldsymbol{\beta}^{[k]})\), the Hessian of the objective function, because the second derivative of the constraint is null. The update problem (\ref{eq:SQP}) is then seen  as a typical quadratic Taylor approximation that can be used to derive descent steps. The linearised constraints ensure that the updated coefficient vector \( \boldsymbol{\beta}^{[k+1]} = \boldsymbol{\beta}^{[k]} + \mathbf{b} \) remains feasible. 

%%% Subsection: CIRLS as a SQP
\subsection*{CIRLS as a SQP}

The CIRLS updating step (\ref{eq:CIRLS2}) can be written as the QP of Equation (\ref{eq:SQP}), where \(f(\boldsymbol{\beta})\) is the minus log-likelihood for an exponential family evaluated at \(\boldsymbol{\beta}\). Using the notations introduced for Algorithm 1, the gradient vector of the log-likelihood can be written as:
%%% Equation: gradient of exponential family
\begin{equation} \label{eq:expgrad}
  \nabla f(\boldsymbol{\beta}^{[k]}) = \mathbf{X}^T\mathbf{W}\mathbf{G}(\mathbf{y} - \boldsymbol{\mu}^{[k]})/\phi
\end{equation}
where $\phi$ is the dispersion parameter related to the exponential family. Note that \(\mathbf{W}\) and \(\mathbf{G}\) depend on \(\mathbf{\mu}^{[k]}\) and will therefore change at each iteration. However, here and in the following, we drop the indexing \([k]\) to simplify the notation.

Second, note that in our case, we only consider linear constraints \(c_r(\boldsymbol{\beta}) = \mathbf{c}_r^T\boldsymbol{\beta}\) with \(\mathbf{c}_r\) the \(r^{th}\) row of the constraint matrix \(\mathbf{C}\). This means that their second derivative is null and that the Hessian of the Lagrangian reduces to the Hessian of the log-likelihood, i.e
%%% Equation: hessian matrix
\begin{equation} \label{eq:exphessian}
  \begin{split}
    \nabla^2 L(\boldsymbol{\beta}^{[k]}) &= \nabla^2 f(\boldsymbol{\beta}^{[k]}) - \nabla^2 \boldsymbol{\lambda} \mathbf{C} \boldsymbol{\beta}^{[k]} \\
    &= \nabla^2 f(\boldsymbol{\beta}^{[k]}) \\
    &= -\mathbf{X}^T \mathbf{W}\mathbf{X} / \phi
  \end{split}
\end{equation}
The full derivation of the Hessian can be found in \citet{Wood2017GeneralizedAdditiveModels}.

The objective function in (\ref{eq:SQP}) being a quadratic Taylor approximation, it means a Newton step for the update \(\mathbf{b} =  -\nabla^2 L(\boldsymbol{\beta}^{[k]})^{-1} \nabla f(\boldsymbol{\beta}^{[k]}) \) provides a descent direction \citep{BoggsTolle1995SequentialQuadraticProgramming}. Adding the Newton step to the current iteration \(\boldsymbol{\beta}^{[k]}\), we find:
%%% Equation: CIRLS update
\begin{equation} \label{eq:CIRLSupdate}
  \begin{split}
    \boldsymbol{\beta}^{[k+1]} &= \boldsymbol{\beta}^{[k]} + \mathbf{b} \\
    &= \boldsymbol{\beta}^{[k]} - \nabla^2 L(\boldsymbol{\beta}^{[k]})^{-1} \nabla f(\boldsymbol{\beta}^{[k]}) \\
    &= \boldsymbol{\beta}^{[k]} + (\mathbf{X}^T \mathbf{W} \mathbf{X})^{-1} \mathbf{X}^T\mathbf{W}\mathbf{G}(\mathbf{y} - \boldsymbol{\mu}^{[k]}) \\
    &= (\mathbf{X}^T \mathbf{W} \mathbf{X})^{-1} \mathbf{X}^T \mathbf{W} \mathbf{X} \boldsymbol{\beta}^{[k]} + (\mathbf{X}^T \mathbf{W} \mathbf{X})^{-1} \mathbf{X}^T \mathbf{W} \mathbf{G} (\mathbf{y} - \boldsymbol{\mu}^{[k]}) \\
    &= (\mathbf{X}^T \mathbf{W} \mathbf{X})^{-1} \mathbf{X}^T \mathbf{W} \mathbf{G}(\mathbf{y} - \boldsymbol{\mu}^{[k]}) + \mathbf{X} \boldsymbol{\beta}^{[k]} \\
    &= (\mathbf{X}^T \mathbf{W} \mathbf{X})^{-1} \mathbf{X}^T \mathbf{W} \mathbf{z}
  \end{split}
\end{equation}
with \(\mathbf{z} = \mathbf{G}(\mathbf{y} - \boldsymbol{\mu}^{[k]}) + \mathbf{X} \boldsymbol{\beta}^{[k]}\). The last equation in (\ref{eq:CIRLSupdate}) is exactly the solution of the least-square objective function \(\left( \mathbf{z} - \mathbf{X}\boldsymbol{\beta} \right)^{T} \mathbf{W} \left( \mathbf{z} - \mathbf{X}\boldsymbol{\beta} \right)\) in the IRLS and CIRLS algorithms (Equation \ref{eq:CIRLS}). This shows the iteratively reweighted least squares are consistent with a SQP iteration. 

Finally, the linearised constraint in (\ref{eq:SQP}) can be rewritten
%%% Equation: constraints
\begin{equation} \label{eq:CIRLSconstraint}
  \begin{split}
    \nabla c_r(\boldsymbol{\beta}^{[k]}) \mathbf{b}^T + c_r(\boldsymbol{\beta}^{[k]}) &= \mathbf{c}_r^T \mathbf{b} + \mathbf{c}_r^T\boldsymbol{\beta}^{[k]} \\
    &= \mathbf{c}_r^T\boldsymbol{\beta}^{[k+1]}
  \end{split}
\end{equation}

which is obtained from the fact that the linear constraints are written \(c_r(\boldsymbol{\beta}^{[k]}) = \mathbf{c}_r^T\boldsymbol{\beta}^{[k]}\) and that the first derivative of such a constraint is simply \(\mathbf{c}_r^T\). The resulting expression in (\ref{eq:CIRLSconstraint}) corresponds to the constraints in the quadratic program in the CIRLS algorithm.

%%% Subsection: convergence properties.
\subsection*{Convergence properties}

SQP are generally known to converge locally even when starting relatively far from the local optimum. Specifically, in our case, both the objective function and constraints are twice differentiable, and the constraint matrix is required to be irreducible. Therefore, the conditions for local convergence are respected \citep{NocedalWright2006NumericalOptimization}. In this case, the algorithm converges quadratically towards the local minimum, which follows from the update being a Newton step \citep{BoggsTolle1995SequentialQuadraticProgramming}. This is consistent with our experience working with the algorithm, in which CIRLS always converges quickly to a solution.

Additionally, when using canonical link functions, the GLM log-likelihood is concave \citep{Wedderburn1976ExistenceUniquenessMaximum}, guaranteeing a unique solution. Therefore, CIRLS is expected to converge to a global optimum within the feasible region. This result is also suggested in previous works \citep{Meyer2013SimpleNewAlgorithm, MeyerWoodroofe2004ConsistentMaximumLikelihood}.

%%%%%%%%%% Section 2: additional details on inference

\section{Coefficients inference}

%%% Distribution
\subsection*{Distribution}

In unconstrained GLMs estimated by IRLS, the coefficient vector \(\boldsymbol{\beta}\) has distribution \citep{Wood2017GeneralizedAdditiveModels}:
%%% Equation: distribution of unconstrained coefficients
\begin{equation} \label{eq:betaDistribution}
  \hat{\boldsymbol{\beta}}^* \sim N(\boldsymbol{\beta}^*, \phi^*(\mathbf{X}^T \mathbf{W}^* \mathbf{X})^{-1})
\end{equation}

where the asterisk in \(\boldsymbol{\beta}^*\), \(\phi^*\) and \(\mathbf{W}^*\) identifies them as from an unconstrained model. To get the distribution of the constrained coefficients, we first apply the affine transformation defined by the constraint matrix \(\mathbf{C}\), and then apply the box constraints given by vectors \(\mathbf{l}\) and \(\mathbf{u}\). This results in a Truncated Multivariate Normal (TMVN) distribution as \citep{Horrace2005ResultsMultivariateTruncated}
%%% Equation: distribution of cconstrained coefficients
\begin{equation} \label{eq:tmvn}
  \mathbf{C}\hat{\boldsymbol{\beta}} \sim TMVN(\mathbf{C}\boldsymbol{\beta}^*, \phi^* \mathbf{C} (\mathbf{X}^T \mathbf{W}^* \mathbf{X})^{-1} \mathbf{C}^T, \mathbf{l}, \mathbf{u})
\end{equation}

from which we can simulate \citep{Botev2017NormalLawLinear, Geweke1991EfficientSimulationMultivariate} and compute moments \citep{ManjunathWilhelm2021MomentsCalculationDoubly}.

%%% Back-transformation
\subsection*{Back-transformation}

To obtain inference for \(\hat{\boldsymbol{\beta}}\), one needs to transform simulated values or moments obtained for the vector \(\mathbf{C}\hat{\boldsymbol{\beta}}\) to the space of \(\boldsymbol{\beta}\). However, this is not easily done if \(\mathbf{C}\) is not square, i.e. if the number of constraints \(m\) is lower than the number of coefficients \(p\). In that case, we augment \(\mathbf{C}\) as
%%% Equation: C augmentation
\begin{equation} \label{eq:Caug}
  \mathbf{A} = \begin{bmatrix} \mathbf{C} \\ \mathbf{H} \end{bmatrix}
\end{equation}
where the rows of \(\mathbf{H}\) are chosen to be orthogonal to those of \(\mathbf{C}\) and orthonormal between themselves \citep{Tallis1965PlaneTruncationNormal}. A natural candidate for \(\mathbf{H}\) is therefore the null space of \(\mathbf{C}^T\). With this augmentation, the bound vectors \(\mathbf{l}\) and \(\mathbf{u}\) also have to be augmented, and we stack vectors of length \(p-m\) containing only \(-\infty\) and \(\infty\) respectively. Therefore, being uncorrelated to \(\mathbf{C}\) and unconstrained, these new variables do not influence the simulation of variables in \(\mathbf{C}\hat{\boldsymbol{\beta}}\) and allow for easy back-transformation.

%%% Reducing the constraint matrix
\subsection*{Constraint matrix reduction}

The inference explained above requires that there are no more constraints than coefficients (\(m \leq p\)) which would prevent inverting \(\mathbf{C}\) and thus back-transforming. When \(m > p\), it is possible to combine TMVN sampling with an accept-reject algorithm where we choose \(p\) constraints to simulate as explained above and only accept draws that respect the remaining \(m-p\) constraints. However, this raises additional challenges related to choosing which of the \(p\) constraints to select. Therefore, it is preferable to remain in a situation in which \(m \leq p\).

It is often possible to reduce \(m\) by removing redundant constraints and detecting underlying equality constraints in \(\mathbf{C}\). As a simple example, assume that one wants to add the constraint that a set of \(p\) coefficients should all be non-negative (i.e. \(\beta_j \geq 0\) for \(j=1,...,p\)) and that they should also be non-decreasing (i.e. \(\beta_j - \beta_{j-1} \geq 0\) for \(j=2,...,p\)). Then the constraints \(\beta_j \geq 0\) for \(j=2,...,p\) are redundant since they can be deduced from \(\beta_1 \geq 0\) and \(\beta_j - \beta_{j-1} \geq 0\) for \(j=2,...,p\). 

%%%% Algorithm for redundant constraints
\begin{algorithm}[h] \label{algo:redundant}
  \caption{Detection of redundant and equality constraints}
  For each constraint \(l_r \leq \mathbf{c}_r^T \boldsymbol{\beta} \leq u_r\)
  \begin{enumerate}
      \item Find the minimum possible value of \(\mathbf{c}_r^T \boldsymbol{\beta}\) allowed by other constraints. This is done by solving the linear program \begin{equation} \label{eq:linp}
        \begin{split}
            \min & \quad c(r)=\mathbf{c}_r^T \boldsymbol{\beta} \\
            \text{s.t.} &  \quad\mathbf{l}_{-r} \leq \mathbf{C}_{-r}^T \boldsymbol{\beta} \leq \mathbf{u}_{-r}
        \end{split}
      \end{equation}
where the index \(-r\) indicates the of constraints with the \(r^{th}\) one removed.
      \item If \(c(r) \geq l_r\) at the solution, it means the constraint is redundant and can be removed by setting \(l_r\) to \(-\infty\).
      \item Repeat for the upper bound, this time replacing the minimisation problem in (\ref{eq:linp}) by a maximisation problem and comparing \(c(r) \leq u_r\).
      \item If we have either \(\min c(r) = u_r\) or \(\max c(r) = l_r\) then there is an underlying equality constraint that can be flagged.
  \end{enumerate}
  Evaluate the underlying equality constraints that have been flagged to reduce the number of constraints.
\end{algorithm}

More generally, consider the constraint \(l_r \leq \mathbf{c}_r^T \boldsymbol{\beta} \leq u_r\), it is redundant if the minimum possible value of \(c(r) = \mathbf{c}_r^T \boldsymbol{\beta}\) allowed by other constraints in the problem is in fact larger than \(l_r\) \citep{Telgen1983IdentifyingRedundantConstraints, Greenberg1996ConsistencyRedundancyImplied}. Furthermore, if the minimum value of \(c_r\) is equal to \(u_r\), then there is an underlying equality constraint, which means further reduction is possible. The same applies to the maximum of \(c(r)\), this time comparing it to \(u_r\) for redundancy and to \(l_r\) for equality. The minimum values of \(c(r)\) under other constraints can be obtained by a linear program (LP) such as the one displayed in Equation \ref{eq:linp} and the maximum by the same program, changing \(\min\) to \(\max\). Algorithm 2 details the full process of constraint reduction that is implemented in the \texttt{cirls} package in R, which is similar to the one of \citet{CaronEtAl1989DegenerateExtremePoint}. This algorithm is computationally efficient, taking advantage of Dantzig's simplex algorithm to solve the multiple LPs.

%%% Distributional properties
\subsection*{Distributional properties}

\citet{ManjunathWilhelm2021MomentsCalculationDoubly} provide explicit formulas for the moments of a TMVN, which can provide information on some properties of the estimator \(\hat{\boldsymbol{\beta}}\). Denoting \(\boldsymbol{\theta} = \mathbf{C}\boldsymbol{\beta}^*\) as the mean vector and \(\boldsymbol{\Sigma} = \phi^* \mathbf{C} (\mathbf{X}^T \mathbf{W}^* \mathbf{X})^{-1} \mathbf{C}^T\) as the covariance matrix in (\ref{eq:tmvn}), the expectation of the \(r^{th}\) element of \(\mathbf{C}\hat{\boldsymbol{\beta}}\) (i.e. \(\mathbf{c}_r\hat{\boldsymbol{\beta}}\) with \(\mathbf{c}_r\) the \(r^{th}\) row of \(\mathbf{C}\))
%%% Equation: TMVN expectation
\begin{equation} \label{eq:TMVNmean}
  \mathbb{E}(\mathbf{c}_r\hat{\boldsymbol{\beta}}) = \theta_r + \sum_{s=1}^m \sigma_{rs} (F_s(l_s) - F_s(u_s))
\end{equation}
with \(\theta_r = c_r \boldsymbol{\beta}^*\), \(\sigma_{rs}\) the row \(r\) and column \(s\) element from the covariance matrix \(\boldsymbol{\Sigma}\), and \(F_s(l_s)\) and \(F_s(u_s )\) the \(s^{th}\) marginal density of the TMVN (\ref{eq:tmvn}) evaluated at its constraint bounds \citep{Cartinhour1990OnedimensionalMarginalDensity}. From (\ref{eq:TMVNmean}), we can see that, unless \(F_s(l_s ) = F_s(u_s)\) for all \(s\), then \(\mathbb{E}(\mathbf{C}\hat{\boldsymbol{\beta}}) \neq \boldsymbol{\theta}\) and so \(\mathbb{E}(\hat{\boldsymbol{\beta}}) \neq \boldsymbol{\beta}^*\). Therefore, the constraints introduce bias in the estimator \(\hat{\boldsymbol{\beta}}\). The formula suggests that \(\hat{\boldsymbol{\beta}}\) is unbiased when \(F_s(l_s ) = F_s(u_s)\) for all \(s\), which happens when all \(\theta_s\) are perfectly centred between \(l_s\) and \(u_s\), that is in the middle of the feasible region.

The covariance matrix of \(\mathbf{C}\hat{\boldsymbol{\beta}}\) is necessarily smaller than \(\boldsymbol{\Sigma}\) since the TMVN is a contraction of a multivariate normal that concentrates the probability mass in a smaller domain. This is shown in the univariate case \citep{BarrSherrill1999MeanVarianceTruncated} and experimentally for the multivariate case \citep{ManjunathWilhelm2021MomentsCalculationDoubly}. The existing formula for the covariance matrix of a TMVN, although complex, shows that the variance is reduced even for untruncated variables if they are correlated with truncated ones. This suggests, for instance, that constraints would also bias and reduce the variance of confounders in epidemiological models as shown in our simulation study. However, conditional independence is preserved, meaning that the variance of untruncated variables that have null correlation with truncated ones are unaffected \citep{KotzEtAl2000ContinuousMultivariateDistributions}. This fact justifies the use of the null space \(\mathbf{H}\) in (\ref{eq:Caug}), with the properties exposed above also being true for \(\mathbf{A}\hat{\boldsymbol{\beta}}\).  

%%%%%%%%%% Section 3: additional details on simulations

\section{Additional details on the simulation study}

\subsection*{Generation of true coefficients}

\begin{figure}[h]
  \includegraphics[width=1\textwidth]{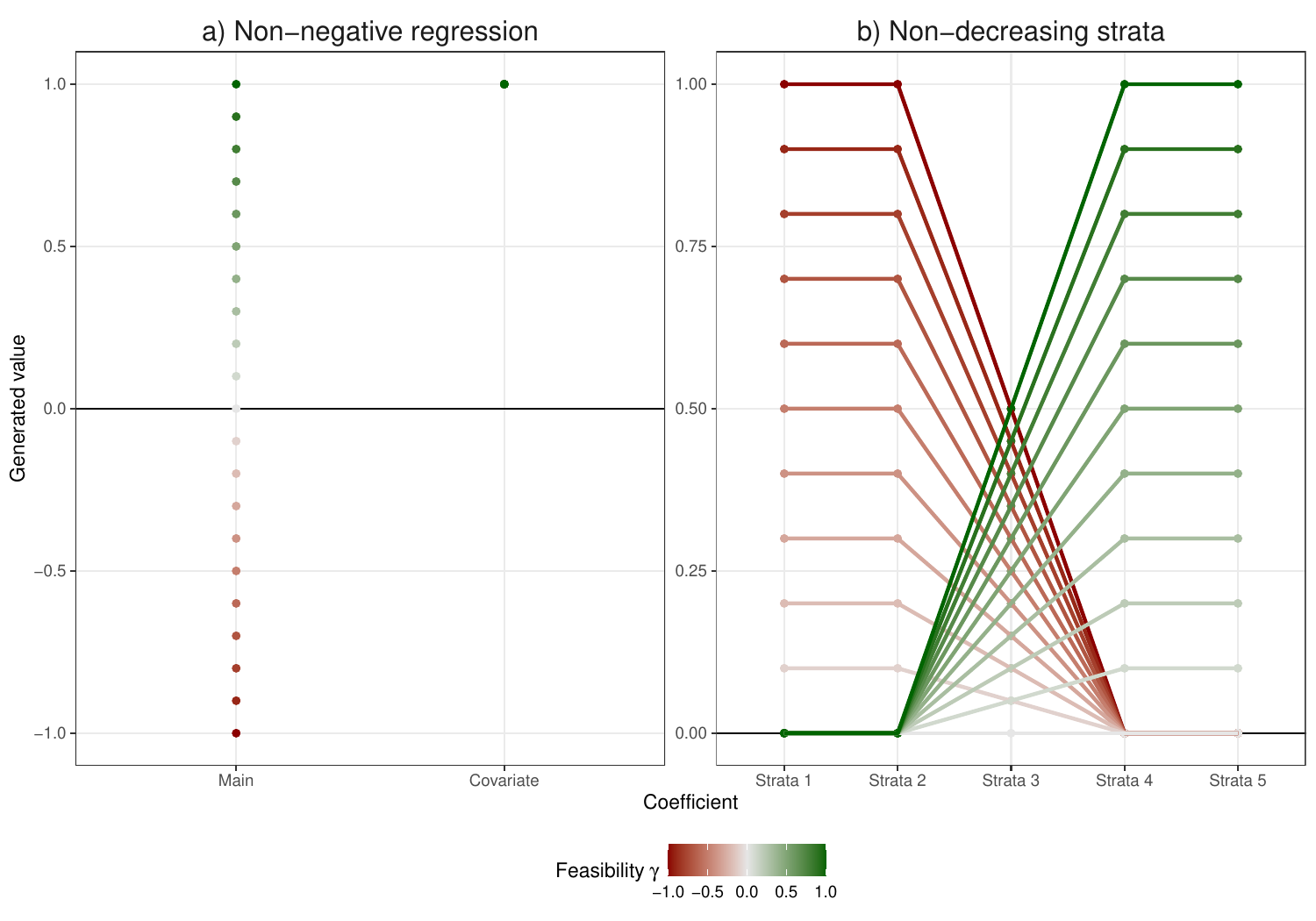}
  \caption{Generated true coefficients from both data-generating mechanisms (DGM) in the simulation study according to the feasibility parameter \(\gamma\). Green coefficients are feasible and red are unfeasible according to the constraints, with white being on the boundary.}
  \label{fig:genCoef}
\end{figure}

Figure \ref{fig:genCoef} shows the true coefficients generated for each value of the feasibility parameter \(\gamma\). In the first DGM, the constraint is \(\beta_1 \geq 0\) and the generated coefficient vector is feasible for non-negative values of \(\beta_1\) (the main coefficient). The covariate coefficient (\(\beta_2\)) is set to 1 and unconstrained, and therefore always feasible. In the second DGM (Figure \ref{fig:genCoef}b), the coefficient vector is feasible when it is increasing and unfeasible when decreasing.

%%% Figure: true generated coefficients

%%% Performance measure formulas
\subsection*{Performance measures}

We detail here the formulae of the performance measures reported in the main manuscript. Figure 1 of the main manuscript shows the difference in several performance measures between the constrained and unconstrained models. For a given coefficient \(\beta_j\) in \(\boldsymbol{\beta}\), these measures are computed as \citep{MorrisEtAl2019UsingSimulationStudies}
%%% Equation: estimation performance measures
\begin{equation} \label{eq:perfmeasures}
  \begin{split}
    \text{Squared bias} &= (\bar{\beta}_j - \beta_j)^2 \\
    \text{Standard error} &= \sqrt{(n_{sim} - 1)^{-1} \sum_i (\hat{\beta}_j^{(i)} - \bar{\beta}_j )^2 } \\
    \text{RMSE} &= \sqrt{n_{sim}^{-1} \sum_i (\hat{\beta}_j^{(i)} - \beta_j )^2 }
  \end{split}
\end{equation}
where \(\hat{\beta}_j^{(i)}\) is the estimated \(\beta_j\) for simulation \(i=1,...,n_{sim}\), and \(\bar{\beta}_j\) is the mean of the \(\hat{\beta}_j^{(i)}\). The Squared Bias measures the systematic deviation of the estimator to the true value, the standard error measures the stability of the estimator across simulations, and the root mean squared error (RMSE) measures the error. Note that we have that \(\text{RMSE}^2=\text{Squared bias}+\text{Standard error}^2\) and minimising the RMSE is therefore a trade-off between bias and standard error.

The performances of the uncertainty assessment are evaluated with:
%%% Equation: estimation performance measures
\begin{equation} \label{eq:infmeasures}
  \begin{split}
    \text{Relative variance error} &= \frac{n_{sim}^{-1} \sum_i \hat{V}(\hat{\beta}_j^{(i)})}{\text{Standard error}^2} \\
    \text{Coverage} &= n_{sim}^{-1} \sum_i \mathbf{1}({\beta_j^{low}}^{(i)} \leq \beta_j \leq {\beta_j^{high}}^{(i)}) \\
    \text{Bias-eliminated coverage} &= n_{sim}^{-1} \sum_i \mathbf{1}({\beta_j^{low}}^{(i)} \leq \bar{\beta}_j \leq {\beta_j^{high}}^{(i)})
  \end{split}
\end{equation}
with \(\hat{V}(\hat{\beta}_j^{(i)})\) the estimated variance from the TMVN for \(\beta_j\), \([{\beta_j^{low}}^{(i)} ; {\beta_j^{high}}^{(i)}]\) is the 95\% confidence intervals estimated for simulation \(i=1,...,n_{sim}\), and \(\mathbf{1}(.)\) the indicator function. The relative variance error, therefore, represents the increase between the estimated variance and the measured variance of the estimated coefficients from the simulations. The coverage counts the proportion of confidence intervals that contain the true value of \(\beta_j\). The bias-eliminated coverage (shown in Figure \ref{fig:becover} below) compensates for under-coverage induced by biased estimators, using the average \(\bar{\beta}_j\) as the reference instead of the true \(\beta_j\).

Observed and expected degrees of freedom are compared to generalised degrees of freedom, computed through the formula \citep{Ye1998MeasuringCorrectingEffects, ShenEtAl2004AdaptiveModelSelection}:%%% Equation: generalised degrees of freedom
\begin{equation} \label{eq:genDF}
  \text{Generalised degrees of freedom} = \phi^{-1} \sum_{i=1}^{n} \text{cov}(g(\hat{\mu}_i), y_i)
\end{equation}
where the terms are defined as in Equation (1) of the main manuscript and \(\phi\) is the dispersion parameter. Formula (\ref{eq:genDF}) can be computed from the simulations as the \(n_{sim}\) replications allow computing the covariance for each observation and \(\phi\) is fixed by the data-generating mechanism. We then report RMSE, bias and standard error between both \(odf\) and \(edf\), and the generalised degrees of freedom.

%%% Figure: bias-eliminated coverage
\begin{figure}[h]
  \includegraphics[width=1\textwidth]{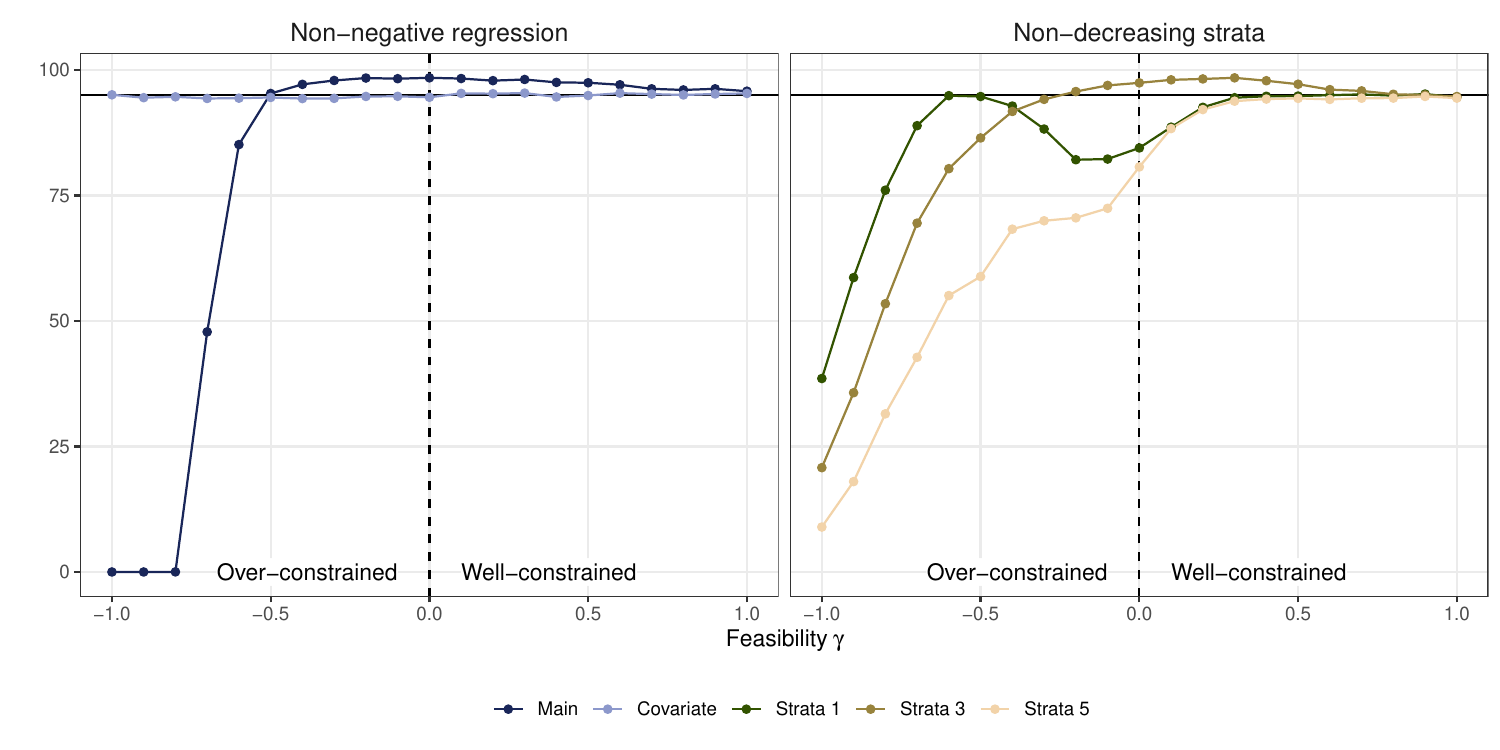}
  \caption{Bias-eliminated coverage for a 95\% confidence interval in the simulation study.}
  \label{fig:becover}
\end{figure} 

%%% Figure: error in DF estimation
\begin{figure}[h]
  \includegraphics[width=1\textwidth]{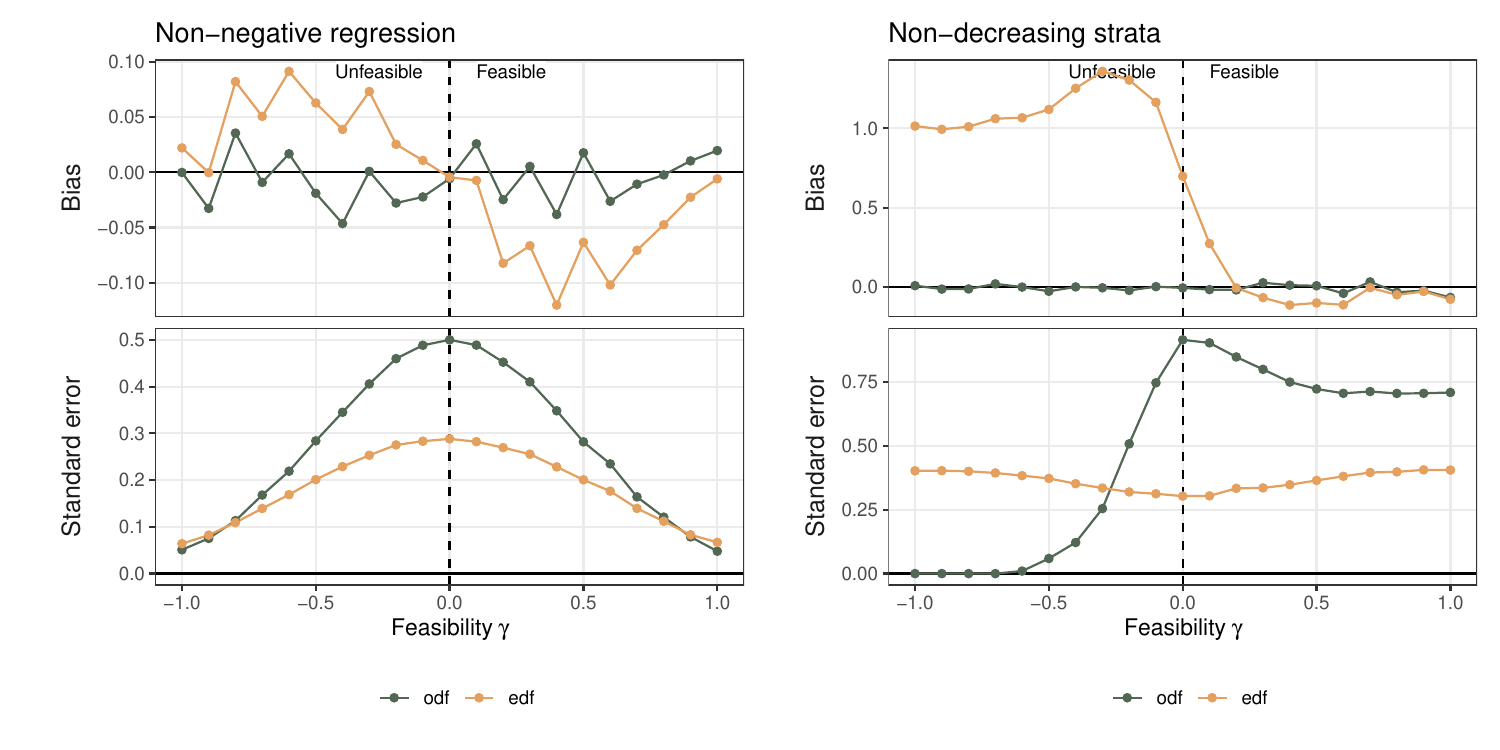}
  \caption{Bias and standard error of odf and edf in estimating simulated degrees of freedom.}
  \label{fig:dferror}
\end{figure} 

%%% Subsection: additional results
\subsection*{Additional results}

Figure \ref{fig:becover} shows the bias-eliminated coverage and shows that the bias is not the only source of coverage error when the true coefficients are not feasible. Figure \ref{fig:dferror} shows bias and standard error in the estimation of degrees of freedom using Ye’s formulae as a comparison. It indicates that \(odf\) is generally unbiased but highly variable while \(edf\) can be biased for highly constrained models but less variable, explaining its lower RMSE in some instances.

\begin{figure}[h]
  \includegraphics[width=1\textwidth]{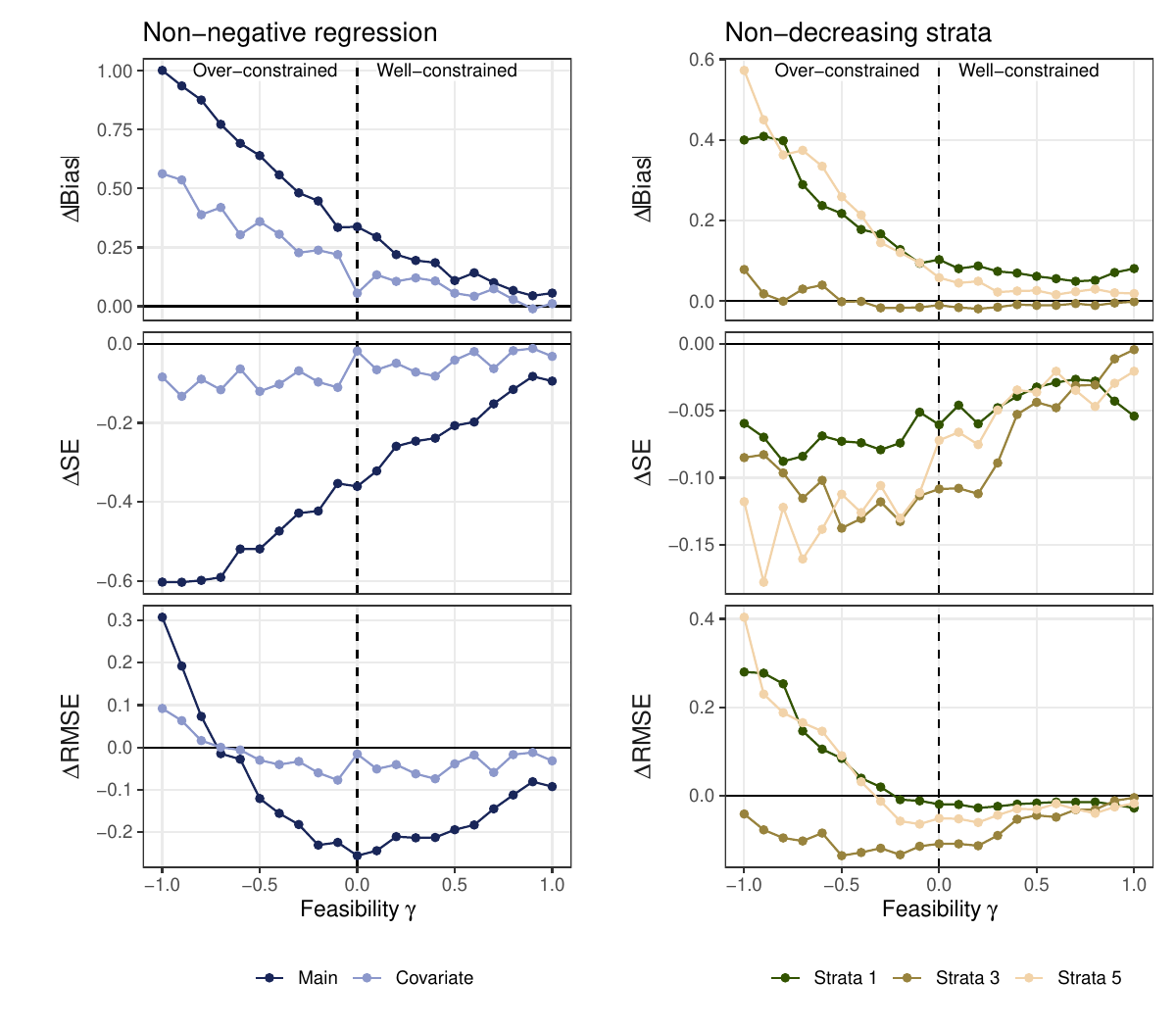}
  \caption{Difference in absolute bias (top row), standard error (middle row) and root mean squared error (RMSE, bottom row) between CIRLS and an unconstrained GLM for the simulation study with \(n=100\). Positive \(\Delta\) indicate lower performances negative \(\Delta\) improved performances of CIRLS compared to the unconstrained estimate.}
  \label{fig:perf100}
\end{figure}

\begin{figure}[h]
  \includegraphics[width=1\textwidth]{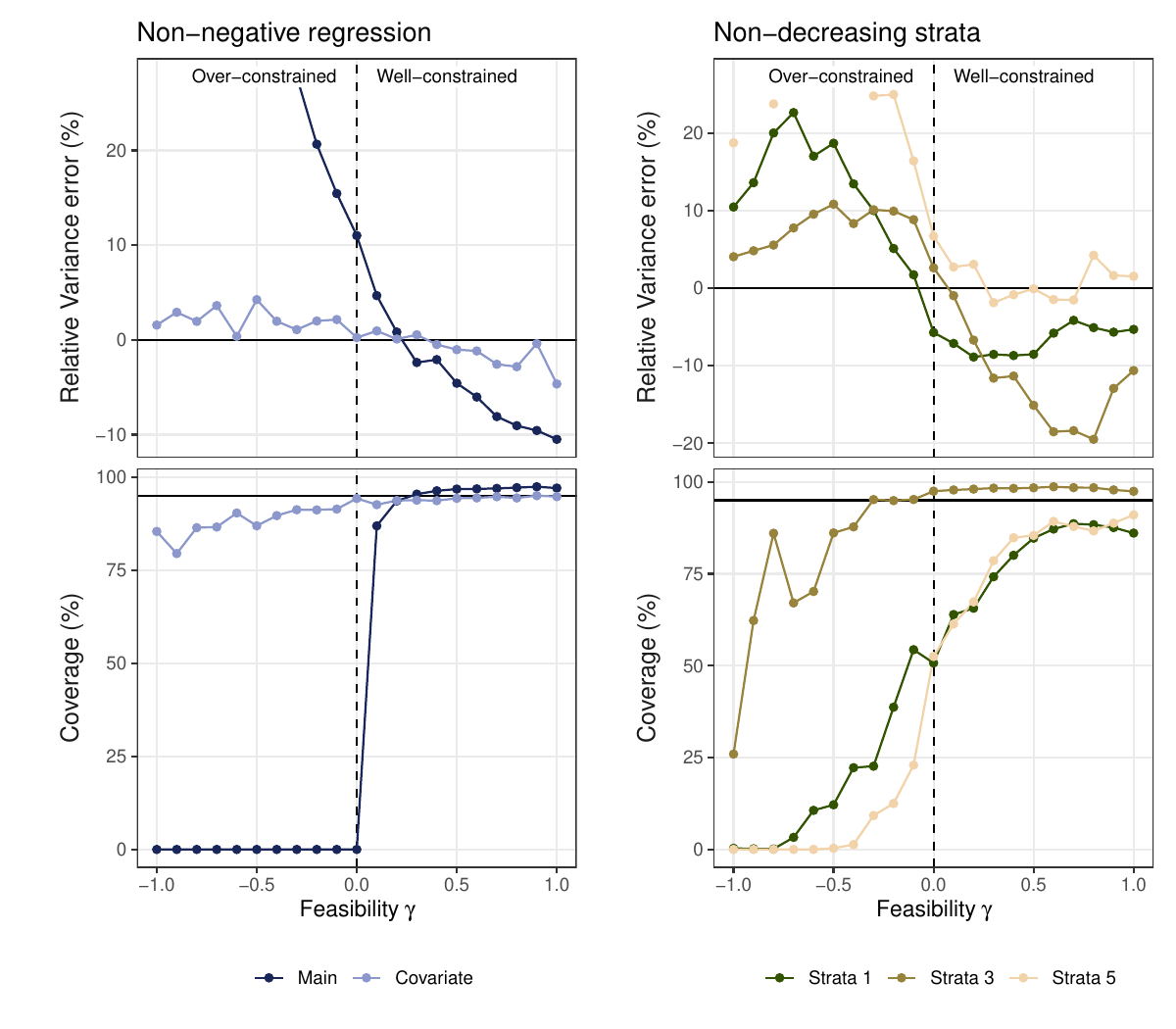}
  \caption{Evaluation of the Truncated Multivariate Normal (TMVN) inference for constrained GLMs versus the feasibility parameter \(\gamma\) for the simulation study with \(n=100\).}
  \label{fig:inf100}
\end{figure}

\begin{figure}[h]
  \includegraphics[width=1\textwidth]{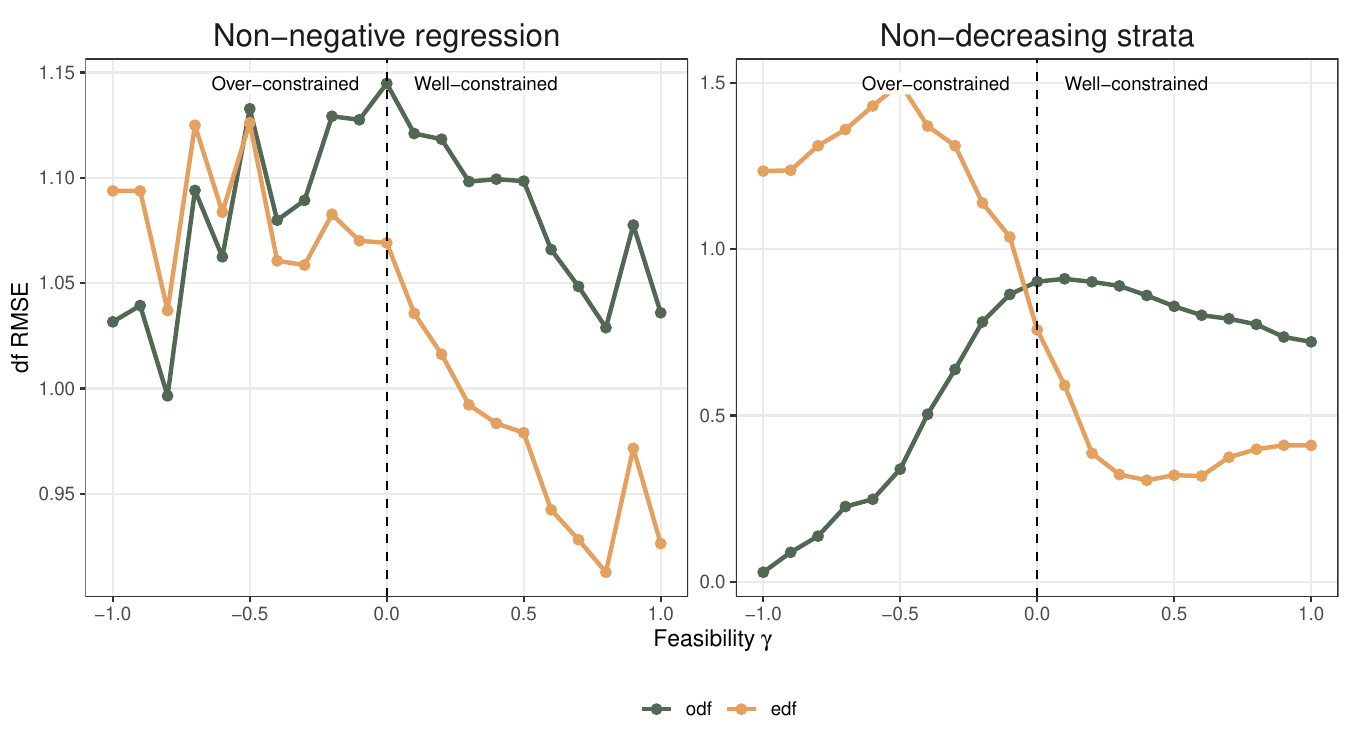}
  \caption{Root mean squared error (RMSE) of observed degrees of freedom (odf) and expected degrees of freedom (edf) in estimating the actual degrees of freedom for the simulation study with \(n = 100\).}
  \label{fig:df100}
\end{figure}

Figures \ref{fig:perf100} to \ref{fig:df100} report the same results as Figures 1 to 3 of the main manuscript but for the simulation study with \(n = 100\). The results are consistent with the results for \(n=500\). In particular, the benefits of CIRLS in terms of RMSE (Figure \ref{fig:perf100}) are extended in this case, with most feasibility parameters \(\gamma\) resulting in negative \(\Delta\text{RMSE}\). This suggests the benefits of CIRLS become even more apparent in very low-power settings.

%%% Figure: bias-eliminated coverage

%%% Figure: error in DF estimation

%%%%%%%%%%%%%%%%%%%%%%%%%%%%%%%%%%
\section{Additional details on the application}
We provide here the full mathematical details of the temperature-related mortality application. The response \(y_i\) follows a Poisson distribution, with rate modelled as:
\begin{equation} \label{eq:tempmod}
    \log(\mathbb{E}(y_i)) = \alpha + \sum_{j=1}^{p_1} \beta_jB_j(x_i) + \sum_{j=1}^{p_2} \theta_jN_j(t_i) + \mathbf{z}_i^T \boldsymbol{\zeta}
\end{equation}
where \(B_j(x_i)\) are basis functions evaluated at temperature \(x_1\) and \(N_j(t_i)\) are natural cubic spline bases evaluated at date \(t_i\). \(N_j(t_i)\) is built to have seven degrees of freedom per year (so 77 in total) with equally spaced knots, which allows to capture the long-term trend and seasonality of daily deaths \(y_i\). The vector \(\mathbf{z}_i\) contains six dummy variables that identify the day of the week. This model specification is typical of time series studies in environmental epidemiology \citep{MasselotGasparrini2025ModellingExtensionsMultilocation}.

In model (\ref{eq:tempmod}), only the \(\beta_j\) coefficients are constrained since this is the association that er are interested about. Therefore, the constraint matrix has the form 
\begin{equation}
    \mathbf{C}=\begin{pmatrix}
        \mathbf{0}_{m\times 1} & \mathbf{C}_{\beta} &\mathbf{0}_{m \times (p_2 + 6)}
    \end{pmatrix}
\end{equation}
where \(\mathbf{0}_{m\times p}\) identifies a matrix of zeros with \(m\) rows and \(p\) columns. The matrix \(\mathbf{C}_{\beta}\) depends on the chosen basis \(B_j(x_i)\).

\subsection*{Binning model}
In the binning model, \(B_j(x_i)\) is a dummy variable taking one when \(x_i\) is in the \(j^{th}\) bin, and zero otherwise. Here, the bins are \(2^\circ\)C wide, with breaks at every even number in the temperature range. Therefore, in this context, \(\mathbf{C}_{\beta}\) is defined exactly as the matrix in (\ref{eq:diffmat}).

Confidence intervals for the coefficients are computed using the standard methods, that is through profiling for the unconstrained GLM, and by computing the \(2.5^{th}\) and \(97.5^{th}\) quantiles of 10 000 TMVN draws for the constrained model.

\subsection*{B-spline model}
Here, the \(B_j(x_i)\) are cubic B-splines with equally spaced knots placed at the quintiles to obtain a smooth function. Therefore, in this model we use \(\mathbf{C}_\beta =  \mathbf{D}^{(2)}\) with \(\mathbf{D}^{(2)}\) as defined for constrained splines in section A above.

For the unconstrained model, confidence intervals are computed through standard methods, that is for a temperature \(x_i\) we have variance \(\sigma_i^2= \mathbf{b}_i^T \mathbf{V}_\beta \mathbf{b}_i\) with \(\mathbf{b}_i^T = [B_1(x_i), ..., B_{p_1}(x_i)] ^T\) and \(\mathbf{V}_\beta\) is the standard (co)-variance matrix of the \(\boldsymbol{\beta}\) coefficients from GLMs. We then have the 95\% confidence interval being \(\pm 1.96 \sigma_i\).

For the constrained model, confidence intervals are constructed by drawing 10 000 samples from the TMVN, and predicting the smooth for each draw. Confidence intervals are then computed as the pointwise \(2.5^{th}\) and \(97.5^{th}\) quantiles to ensure the bands also follow the convexity constraint.

%%%%%%%%%%%%%%%%%%%%%%%%%%%%%%
% End

%%% References if bibTeX is used
%%%
%%% Please, do not specify any \bibliographystyle{} command!
%%%
%%% It is already specified in the smj.cls and its
%%% second specification here causes error.
%%% ------------------------------------------------------------
\bibliography{bibliography}

\end{document}